\documentclass[aps,pra,twocolumn,groupedaddress,showpacs]{revtex4}
\usepackage{graphicx}
\usepackage{epstopdf}
\usepackage[usenames,dvipsnames]{color}
\begin{document}

\title{\bf The parametric spring--mass system, its connection with non-linear optics, and an approach for undergraduate students}
\author{I. Boscolo$^a$, F. Castelli$^b$ M. Stellato$^b$} 
\author{S. Vercellati$^a$} 
\email{stefano.vercellati@uniud.it}
\affiliation{$^a$ Department of Chemistry Physics and Environment,
Universit\'{a} degli Studi di Udine, via delle Scienze 208, 33100, Italy\\
$^b$ Department of Physics, Universit\'{a} degli Studi di Milano, via Celoria 16, 20133, Italy}

\date{\today}

\begin{abstract}

The spring--mass system studied in undergraduate physics laboratories may exhibit complex dynamics due to the simultaneous action of gravitational and elastic forces in addition to air friction. In the first part of this paper, we describe a laboratory experiment aimed at beginner students which also gives those with a more advanced background an opportunity to explore more complex aspects of the motions involved. If students are not given predefined apparatus but are allowed instead to design their own set-up for the experiment, they may also learn something about the thought processes and experimental procedures used in physics. In the second part of this paper, we present a systematic study of the parametric behavior of the system because teachers have to master its dynamics. The non-linear interaction between the vertical and the pendular oscillations in a vertical spring--mass system depends on the ratio between the frequencies of the two motions and on the motion's excitation. Systematic experimental investigations, coupled with relevant simulations, highlight the many aspects of physics involved in the partition of energy transfer between the two modes. The different motion waveforms obtained by sweeping through the resonance curve and by applying small and strong excitations are presented and analyzed. The influence of the unavoidable spurious motions is investigated.  An analogy between the parametric interaction in  a spring--mass system and that found during frequency conversion in non-linear optical crystals is discussed. 

   \end{abstract}



\maketitle
\section{Introduction}

The spring--mass system is a common and easy to perform experiment often used to introduce the study of simple harmonic motions during the first academic year of graduation in physics \cite{ib-lajpe}.

Although the experimental apparatus is simple, an actual spring--mass system behaves as a simple harmonic oscillator only under specific conditions: (i) the mass of the spring must be negligible compared to the attached mass; (ii) the spring elongation caused by the attached mass must exceed the spring equilibrium length by one-fourth \cite{olsson}; and (iii) the initial spring stretch must be strictly vertical.

\begin{figure}[!ht]
\centering
\includegraphics[width=4.5cm]{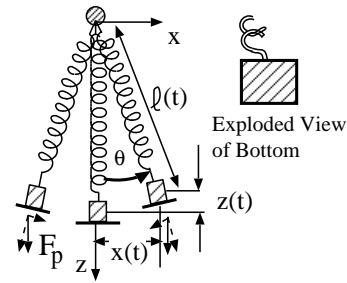}
\caption{\label{forze-moti} Diagram of the applied force with a coupling between the vertical spring and the horizontal
pendular oscillations.}
\end{figure}

If these conditions are not satisfied, the resulting motion of the attached mass will not be a simple harmonic one. It will pass from a vertical to a horizontal oscillation in an apparently random manner, exhibiting parametric rather than harmonic oscillations \cite{olsson,cayton,christ,geba,gallo,arms,cush}. 
A parametric oscillator is a harmonic oscillator that has at least one parameter oscillating over time. In the spring--mass system, the pendular motion of the mass modulates the component of the gravitational force, acting along the axis of the spring (Fig. \ref{forze-moti}).

This modulated component of the gravitational force in combination with the elastic force exerted by the spring generates parametric instability \cite{olsson,cayton,geba,cush} and multimode operation which drives the system at specific frequencies and generates energy transfer between the two modes \cite{cush}, namely the spring-bouncing mode and the pendulum-swinging mode, which can be simultaneously active.
 
The amplitude of the spring oscillation decreases while the amplitude of pendular oscillation increases, and vice versa. There is a coupling between the spring and pendulum oscillations, which is at a maximum when the spring oscillation motion's frequency is double that of the pendular swinging motion.

Using standard symbols, where $\omega_k$ is the oscillating spring's frequency and $\omega_p$ is the oscillation pendulum's frequency, we can express the Frequency Ratio (FR) between the two frequencies as:
\begin{equation}
\label{ratio}
FR=\frac{\omega_k}{\omega_p} = \frac{\sqrt{k/m}}{\sqrt{g/\ell_z}} =2.
\end{equation} 
This is called \textit{resonance condition}. The strong coupling between the two modes induces  parametric instability. This instability is observed within an FR interval around the resonance frequency, with a complete mode exchange near to the center and a partial mode exchange outside of it. The resonance width depends on the degree of motion disorder. That is, the amplitude of the spurious motions (mass wobbling, the transverse vibration of the spring, and the rotation around the spring axis) which are unavoidable in an actual system.

In a laboratory context in which students are allowed to choose and set-up their own experimental apparatus using different springs and masses (which we call {\em open lab}), students can reach particular configurations which are close to the one in which the proper oscillation of the spring elongation and the pendulum oscillation have a ratio not far from two. Students have to approach the spring--mass experiment with an open mind, dealing with the gap between theory and practice, examining complex multi-effect motions, and mastering experimental techniques. Moreover, they have to handle data statistically and purge the experiment of many nuisances. It requires students to develop the ability to manage unexpected experimental observations and perform research-like investigation activities in order to interpret the observed motions. In fact, the analysis of phenomena (which are not fully understood in advance) requires the development of strategies for identifying the motion components, and separating their inter-relations so as to treat each component separately. Afterwards, these components can be combined in order to understand the overall motion. 


\section{The experimental set-up}

\begin{figure}[!ht]
\centering
\includegraphics[width=5cm]{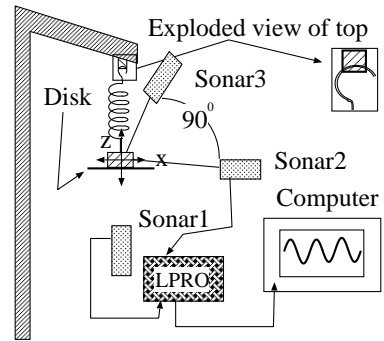}
\caption{\label{setup} Experimental set-up: Sonars (ultrasonic motion detectors) detect the motion. The digital data acquisition tool LPRO - a Vernier Loggerpro - acquires and sends the mass coordinates $z(t)$, $x(t)$ to the computer. The first coil of the spring is fixed to the top rod.}
\end{figure}

The experimental apparatus was kept as simple as possible: a spring and a weight oscillating vertically, as shown in Fig. \ref{setup}.
The $z(t)$ and $x(t)$ coordinates of the hanging mass, defined in relation to its equilibrium position, are tracked respectively by a ultrasonic motion detector located underneath (Sonar1) and  two side ones (Sonar2 and Sonar3).

The bob of the mass-spring system is composed of a stack of 20g metal disks placed on a support to which plasticine may be added in order to obtain the precise desired weight. Below the support, there is a 70mm diameter diskette which optimally reflects the sound waves emitted by the sonar underneath. This expedient is necessary because the mass has wide lateral oscillations and wobbles around its pivot. Another spurious motion can be observed, namely the transverse spring vibration which is due to the mass wobbling. The diskette diameter is a compromise between a larger size for ideal sonar detection, and a smaller size to minimize both friction and spurious motion. These haphazard motions, superimposed on the bouncing and pendular oscillations, cause the observed motion waveforms to be irregular. Distorted waveforms prevent us from attempting a simple analysis of the experiment and its physics. To minimize spurious motion, it is essential to start with a mass displacement which is smooth and precise, and to use small oscillation amplitudes.

\begin{figure}[!ht]
\centering
\includegraphics[width=7.5cm]{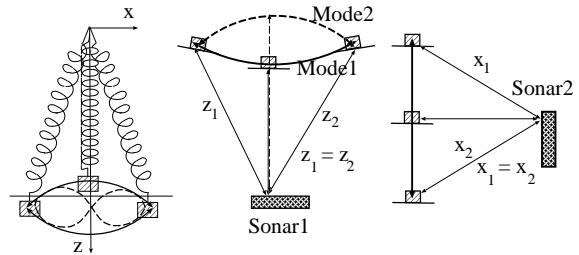}
\caption{\label{view-sonars-moti} Outline of the sound wave paths between the sonars and the appended mass along their trajectories. The convex, concave and Lissajous trajectories represent the motion configurations of the parametric system.}
\end{figure}

\begin{figure}[!ht]
\centering
\includegraphics[width=6cm]{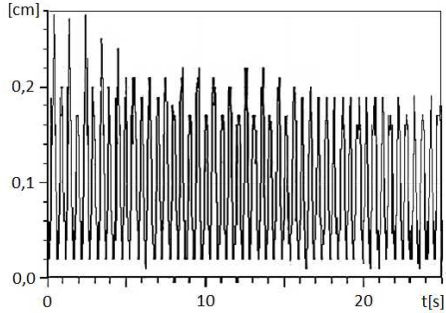}
\caption{\label{Sonar1-pendolo} The waveform detected by Sonar1 (which is located underneath) during pure pendulum oscillation. This should be subtracted from the composite motion waveform detected by Sonar1.}
\end{figure}

Sonar1 and Sonar2 mainly detect the motions in front of them (vertical and transverse), but they also faintly detect longitudinal motions (transverse and vertical respectively) because of the different distances between the sonar and the mass. Sonar1 also partially detects transverse pendular oscillation, as shown in Fig. \ref{view-sonars-moti}. 
Similarly, Sonar2 partially detects vertical spring oscillation. In particular, Sonar2 detects the vertical oscillation amplitude with a reduction factor of more than ten (hence negligible), while Sonar1 detects transverse pendular oscillation with a reduction factor of only about three. A waveform of this detection is shown in Fig. \ref{Sonar1-pendolo} (the mass was appended to a wire for this measurement). This waveform is superimposed on the actual motion waveform, thus introducing a noticeable distortion.
\\
The mass trajectories are mostly convex and concave arcs, as shown in Fig. \ref{view-sonars-moti}. When the trajectory is a convex arc, the vertical oscillation seen by Sonar1 is greatly reduced. Hence, the distorting waveform subtraction technique is required in these cases in order to obtain the waveform that reproduces the actual motion observed by the naked eye.

The detection of the vertical motion by Sonar1 is not affected by the rotation of the pendular oscillation plane. This is not the case for Sonar2. In order to effectively detect the pendular motion in any plane, a third detector, Sonar3, has been set perpendicular to Sonar2, see Fig. \ref{setup}.  The top coil of the spring is fixed to the bar to partially constrain the rotation of the pendulum oscillation plane. Additionally, it must be pointed out that the cylindrical shape of the appended mass renders the vertical motion within the length of the cylinder undetectable, i.e. half a centimeter.

The typical values of mass, spring constant and system length used were $m_{spring}\simeq 4.8 \; g$, $k\simeq 8
\; N/m$, $\ell \sim 21\, cm$ respectively.  

{\em Note}. The apparatus planned for the class experiment uses only Sonar1. Students do not elaborate on the complex motion seen during detection.


\section{The class approach in the context of the open lab}

The spring--mass experiment in the first year of an undergraduate course is intended to introduce students to the physics of harmonic oscillation. The classical harmonic motion is represented by the following equation \cite{halliday}:
\begin{equation}
\label{z(t)-freemotion}
z(t) \,  =  \, z_0 \, e^{-\gamma\, t} \, \cos(\omega\, t),
\end{equation}
where $z_0$ is the initial oscillation amplitude,
$\gamma=C/2m$ is the damping constant, C the damping coefficient, and
\begin{equation}
\label{eq-II}
 \omega^2 = \frac{k}{m} - \left(\frac{C}{2 m}\right)^2
= \omega_k^2  -  \gamma^2  .
\end{equation}

In these equations, the mass of the spring is assumed to be zero, which is not true in an actual apparatus. The damping force is assumed to be proportional to velocity, although air friction against a moving object depends on its shape, and it is not linear with velocity. The experimental investigation involves the following steps: (a) testing Hooke's law $F = -k\, \Delta z$, (b) testing $\omega^2 = k
/m$, (c) measuring the damping time $\tau = 1/\gamma$, and (d)
testing the sinusoidal solution.
 
In an ordinary lab, this experiment is carried out by guiding students to choose heavy masses aimed at obtaining harmonic oscillations. In the context of the open lab approach, students set-up their own experimental apparatus by choosing springs and masses which leads to some of the oscillators having very irregular motions. Hence, the new phenomenon of parametric oscillations comes to the fore. In the class discussion, students understand that the system exhibits the expected harmonic behavior only with heavy masses. The experiment is initially carried out by studying harmonic oscillations with heavy masses, while investigation of the irregular motion is addressed at the end.


\subsection{Measurement of $k$, $\omega$, $\tau$ by meterstick and stopwatch}

The testing of Hooke's law  $F = -k\, \Delta z$ is performed by measuring spring elongations relative to different masses.

The testing of the physical law $\omega^2 = k /m$ is conducted by measuring the periods $T$ relative  to different masses. The range of masses that can be used cannot, however, exceed certain values - they cannot be too large in order to avoid damage to the spring, and they cannot be too little in order to avoid disordered system vibrations. The law can be rewritten in terms of the measured quantities $(T,m)$
\begin{equation}
\label{m-T-eq}
 m  =  \frac{k}{4 \pi^2} T^2.
 \label{emme}
\end{equation}

A graph of the attached masses versus $T^2$ shows the anticipated straight line, but with a small negative intercept on the $y$-axis. The class discussion led to the conclusion that the mass of the spring must be taken into account \cite{ib-lajpe}. An effective mass, $m_e = m +m_s$, operates in the system where $m_s$ is the contribution of the mass of the spring. Hence,
Eq. \ref{emme} must be rewritten as
\begin{equation}
\label{eq-mass}
m  =  \frac{k}{4\pi^2} T^2 - m_s .
\end{equation}
The $y$-intercept of the data graph shows the expected value of one-third of the total spring--mass ($ m_s = m_\mathrm{spring}/3$), as reported in the literature \cite{christ,halliday}. For the remainder of this paper, system mass ($m$) refers to the effective mass as defined above.

While students noticed that lighter masses caused more complex oscillations, they also realized that the system did not behave as a simple harmonic oscillator, and therefore questioned the teacher for an effective explanation. They were advised to delay investigating the non-harmonic motion until the final step of the lab work.

At this point, the students measured the decay time $\tau = \gamma^{-1}$ , that is, the time interval between the beginning of the oscillation and the moment when the amplitude is reduced to $e^{-1}$ of its initial value. The students, looking for reasonably stable oscillations, used large masses in this measurement. 
The expected mass dependence $\tau=(2/C) \cdot m$ was confirmed. However, the results were scattered. The class discussion on the scattered results led to the conclusion that the decay was dependent on the oscillation amplitude \cite{ib-lajpe}.


\subsection{The motion analysis via the waveform}
\label{stau}

\begin{figure}[!ht]
{\includegraphics[width=8cm]{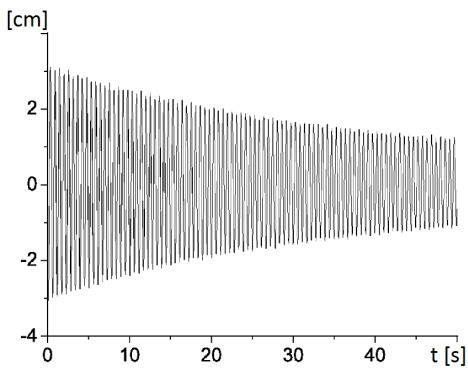}}
\caption{\label{damped-oscill} This waveform is obtained with 60g mass and a very careful vertical stretch.}
\end{figure}

The study of the motion was carried out, from now on, by detecting waveforms and extracting the frequency $\omega$ and the decay time $\tau$ from the waveforms. The waveforms detected by Sonar1 were displayed on the computer screen. A typical waveform is shown in Fig. \ref{damped-oscill}.

The value of the oscillation frequency is obtained by measuring the waveform's period. Subsequently, the Fast Fourier Transform is applied. In this context, students are introduced to the concept and use of the FFT.

The value of the waveform's decay time $\tau$ is obtained by measuring the time interval from the beginning of the oscillation $z_0$ to the $z_0/e$ point.
After the class discussion, students realized that measuring $\tau$ could be done by investigating any section of the curve. However, different curve sections provide different $\tau$ values \cite{ib-lajpe}.
Indeed, the decay time turns out to be non-exponential soon after the beginning of the curve and becomes much longer along the curve. This indicates that the damping coefficient decreases as the oscillation amplitudes decrease; that is, the effect of air friction (against the disk holding the mass) decreases as velocity decreases. As a result, the mathematical form of Eq. \ref{z(t)-freemotion}, based on a constant decay value, does not reproduce actual, observed motion.


\subsection{The unexpected parametric motion during the class experiment}

Once the mandatory testing and measuring of the harmonic motion characteristics  were completed, students looked at the waveforms that were obtained with the use of lighter masses. The complex motion instability which displayed an interchange between the vertical and transverse oscillations made it clear that a new physical model was needed to account for the unexpected behavior. The system set-ups, using masses which ranged from heavy (70g) to light (40g), generated waveforms ranging from decaying exponential sinewaves (as shown in Fig. \ref{damped-oscill}) to partially modulated decaying sinewaves, to fully modulated (multi-lobed) decaying sinewaves. A detailed analysis of the motion is presented in the remainder of this paper. This result leads us to conclude that there are two different classes of motion: harmonic and non-harmonic. The two types of motions refer to different physical phenomena: a harmonic oscillator and an unstable parametric oscillator. The class discussion introduced students to the basics of the subject of parametric instability.


\section{The systematic study of parametric physics}

In order to perform a systematic study of the parametric behavior of a spring--mass system, we have investigated the motion at many Frequency Ratio (FR) values distributed within the resonance curve centered around $FR=2$. At each FR configuration, the set of excitations reported in Table \ref{excitations} were applied.

\begin{center}
\begin{table}[!ht]
\begin{tabular}{|l|l|}

\hline
vse/tse  & vertical-transverse small excitation, 1--2 cm \\
\hline
vle/tle  & vertical-transverse large excitation, 2--3 cm \\
\hline
vse/vle+s  & vertical small/large excitation + seed\\
\hline
tse/tle+s  & transverse small/large excitation + seed\\
\hline

\end{tabular}
\caption{\label{excitations} A set of excitations turning the system in motion. In the acronym vse/vle+s excitation, the term \textit{seed}, is intended as a tiny sideways mass shift added to the vertical shift (spring stretching). In the acronym tse/tle+s excitation, the opposite happens, a small downward shift is added.}
\end{table}
\end{center}

The set of masses used to change the FR, whilst maintaining the same spring, is listed in Table \ref{masses}.

\begin{center}
\begin{table}[!ht]
\begin{tabular}{|c|c|c|c|c|c|c|c|}
\hline
mass weight [g] & 40 & 45 & 50 & 55 & 60 & 70 & 80 \\ 
\hline
FR   & 2.07 & 1.99 & 1.92 & 1.85 & 1.78 & 1.67 & 1.6 \\
\hline
\end{tabular}
\caption{\label{masses} Masses and their respective frequency ratio FR.}
\end{table}
\end{center}

The objective of this study was to find the width of the resonance curve, and thus the efficiency of the parametric generation at different positions within the resonance curve, the threshold of the instability onset, and the dependence of the parametric process on the kind of excitation.
\\When applying a proper excitation, the system (when out of resonance) exhibits an exchange of energy from the spring mode to the pendular mode. This observation suggests an analogy between this behavior and the parametric generation in non-linear optics, as described next.

Equations describing the mode exchange presented in the literature \cite{olsson, cayton} had to be revisited in order to obtain good simulations with the experimental waveforms.
In Fig. \ref{setup}, the graphs of the motions detected by Sonar1 and Sonar2 are referred to as waveform1 and waveform2, and the relative spectra are referred to as spectrum1 and spectrum2 respectively.

\section{The motion equations}

Let us consider a system with a massless linear spring suspended from a rigid point and terminated with a point mass  $m$. Assuming that the motion occurs in a plane, let us define a polar reference frame with its origin at the suspension point: $\ell(t)$ the length of the spring, and $\theta(t)$ the angle of oscillation. For ease of calculation, we define the following quantities:

\begin{displaymath}
\begin{array}{ll}
 \ell_0 =  \textrm{the unloaded spring length} \\
\ell_m =(mg/k) = \textrm{spring elongation by the mass at $\theta=0$} \\
\ell_z = \ell_0+\ell_m  \\
\delta\ell(t)  =
\textrm{extra elongation by spring stretching.}    
\end{array}
\end{displaymath}

Since $\ell_m cos \, \theta$ is the spring elongation by the appended mass at the angle $\theta$, the general length of the spring at the angle $\theta(t)$ is
\begin{equation}
\ell(t) = \ell_0 +\ell_m\, \cos\, \theta(t) + \delta \ell(t)
\end{equation}
The potential $U$ and kinetic $T$ energies in terms of $\ell$ and $\theta$ coordinates are

\begin{eqnarray}
U &=& U_{elastic} + U_{gravitational} = \nonumber \\ 
& = &\frac{1}{2} \, k (\ell_m \, cos \, \theta + \delta \ell)^2 - mg \, \ell(t)\, cos\, \theta = \nonumber \\ 
& = &\frac{1}{2} \, k \,\delta \ell^2 - \frac{1}{2} \, k \, 
\ell_m^2 cos^2 \theta -  k\, \ell_m \,  \ell_0  \cdot cos\,\theta
\end{eqnarray}
\noindent and 
\begin{eqnarray}
\label{T-energy}
T & = &\frac{1}{2} \, m\, [\dot{\delta \ell^2} - 2\, \ell_m \dot{\delta \ell} sin\ \theta \cdot \dot{\theta} + \ell_0^2 \cdot \dot{\theta}^2 +  \delta \ell^2 \cdot \dot{\theta}^2 +    
\nonumber \\
& + & \ell_m^2 \cdot \dot{\theta}^2+ +2\, \ell_0 \, \delta \ell \cdot \dot{\theta}^2 +  2\, \ell_0 \, \ell_m \, cos\, \theta \cdot \dot{\theta}^2 +
\nonumber \\
& + & 2 \, \ell_m \, \delta \ell \, cos\, \theta \cdot \dot{\theta}^2] 
\end{eqnarray}

The two non-linearly coupled Euler–Lagrange equations of motion result work out as:
\begin{equation}
\label{eq1-parametric}
\ddot{\delta \ell} + \omega_k^2 \, \delta\ell = (\ell + \ell_m\, cos\, \theta) \cdot \dot{\theta}^2 + 
\ell_m\, sin\ \theta \cdot \ddot{\theta} 
\end{equation}
\begin{eqnarray}
\label{eq2-parametric}
 \ell^2 \, \ddot{\theta} + g \ell \, sin \theta & = & 
- \ell_m^2 sin^2\theta \cdot \ddot{\theta} + g \, \delta \ell \, sin \theta -  2 \, \ell \, \dot{\delta \ell} \, \dot{\theta} + \\
\nonumber
&&
+ \, \ell_m \, [\ell - \ell_m cos \theta] sin\, \theta \cdot \dot{\theta}^2 + \ell_m sin\ \theta \cdot \ddot{\delta \ell} \\
\nonumber
\end{eqnarray}
It should be stressed that $\ell$ depends on the two $\theta$ and $\delta\ell$ variables.

During the measurement, the typical actual motion had a rotation angle $\theta$ of about 0.1 rad and a spring elongation $\delta \ell$ of 3 cm, which corresponded to roughly $0.13\, \ell$. These rotation and elongation values were applied in the experiment in order to overcome the effect of the spurious motions. The approximation of small angle and small elongation is not applicable in an actual system. However, it is useful for a physics investigation looking at approximate equations. Applying the following approximations,

\[cos \theta =1 -\frac{1}{2} \,\theta^2;\,   sin \theta =\theta; \,
 cos \, \theta \, sin \theta \simeq \theta;\]
 
\[cos^2 \theta = 1- \theta^2;\, \theta^2<< 1;\, \delta \ell << \ell_0\]

\noindent we get the approximate final equations of motion:
\begin{eqnarray}
\label{eq1-approx}
\ddot{\delta \ell} + \omega_k^2 \,\delta \ell &=& (\ell_0 + 
2\, \ell_m\ ) \cdot \dot{\theta}^2 + 
\ell_m\, \cdot \theta \cdot \ddot{\theta}
\\
\label{eq2-approx}
\ddot{\theta}  + \omega_p^2\, \theta & = &
-  \frac{2}{\ell_z} \cdot (\dot{\theta} \,\dot{\delta \ell})
+ \frac{\ell_m}{ \ell_z^2}\, (\ell_0 \, \dot{\theta}^2+ \ddot{\delta \ell}) 
\cdot  \theta.   
\end{eqnarray}

It is evident from the coupling coefficients that the expected motion is very complex, and therefore difficult to predict. Since these motion equations show that the modes are not coupled at the lowest order, the onset of the instability at resonance is expected after a certain excitation intensity. What happens in the region around resonance is a matter for investigation. From the non-linear down- and up-conversion picture, we should instead expect no threshold for frequency generation when a seed is injected into the non-linear system. We should expect a transition time before the onset of a steady state. In Ref.\cite{cayton}, for an ideal system, it is reported that at the start of the growing solution, there is a phase arrangement time up to the settlement of $\pi$ radians of the relative phase between the two modes.

Eqs. \ref{eq1-parametric} and \ref{eq2-parametric} are solved numerically and then converted from polar to Cartesian coordinates for a comparison between the simulations' and the experiments' results. An ad hoc friction term is added into the equation to reproduce the actual friction effect.

\section{A common model of nonlinear interaction  between modes 
in spring--mass  and interaction of waves in nonlinear media}

In the vertical spring--mass system, the pendulum's oscillation modulates the spring's elongation (via the term $mg\cos\theta$) and, conversely, the spring's oscillation modulates the pendulum's length. These two modes are coupled when their frequencies are within a certain interval (the resonance interval), whereas they are de-coupled when the two frequencies are out of resonance. When the modes are coupled, they exchange energy. That is, one mode grows and decays alternately with the other mode (in parametric terminology, the oscillations grow and decay via a parametric instability), as shown by the experiments presented next.

In non-linear media (either plasma or crystal) an incoming wave, the pump wave ($\omega_3,k_3$), modulates the nonlinear polarization $\vec{P}$, and, when the “resonance” condition is met, a pair of waves, ($\omega_1,k_1$) and ($\omega_2,k_2$), are generated whose frequencies and wave numbers are
\begin{equation}
\label{frequencies-sum1}
\omega_3=\omega_1 \pm \omega_2 
\end{equation}
and
\begin{equation}
\label{wavenumer-eq}
\vec{k_3} = \vec{k_1} + \vec{k_2}.
\end{equation} 
The process of energy transfer from the pump wave to the subsidiary waves continues until the pump wave is completely depleted, and then the process reverses. In optics, non-linear parametric interaction is an important phenomenon for the generation/amplification of coherent radiation of a higher or lower frequency than that of the pump wave \cite{jackson,svelto,yariv}. It is used in the optical parametric oscillator (OPO) laser scheme to provide a tunable source of coherent radiation, and also in a non-linear converting crystal, that is, a non-linear crystal crossed by the three waves \cite{aegis}.

While transferring energy from one mode to another, the spring--mass parametric oscillator can be thought of as a frequency converter. In this model, the earlier frequency rule has to hold, the so-called idler frequency $\omega_i$ so that
\begin{equation}
\label{frequencies-sum}
\omega_k=\omega_p+\omega_i
\end{equation}
has to be experimentally observed. In fact, the spectra obtained display idler lines. Different kinds of interplay between the waves, depending on the initial wave intensity, have been reported \cite{yariv}. The same is expected with a different division of energy between the two modes.

\section{Experimental investigation of the system's motions}

With a fixed spring constant, a system configuration is defined by the pendulum length and by the attached mass. These two parameters determine the pendulum and spring frequencies respectively. The ratio between these two frequencies, FR, is the fundamental system parameter.

A system's behavior depends on its configuration and on the applied excitation. The systematic experimental investigation of the parametric behavior requires the identification of a set of relevant configurations and the application of the set of different excitations in Table \ref{excitations} to each configuration. 

A large excitation will drive the system into strong non-linearity, while a small excitation will drive it to a relatively low non-linearity. In order to investigate how the variation of the non-linear coupling changes the system's motion, small and large spring elongations and small and large pendulum shifts (vse, vle, tse, tle) were applied.

On the basis of an analogy with non-linear optics, the concept of the input signal amplification by a pump wave was transferred to this system by applying a very small transverse shift (which we call \textit{seed}) in addition to the vertical excitation (vse+s and vle+s). The seed simulates the optical input signal, the vertical excitation simulates the pump wave.

Three representative configurations are presented and discussed: FR 1.85, 1.92 and 1,99; i.e. far from resonance, near resonance and at resonance. The waveforms are reported and discussed in comparison with the simulations, and by observing the direct motions. Only the two spectra relative to the mid-configuration are reported since they are typical of all the other configurations.

\subsection{Behavior with vse excitation}

\begin{figure}[!ht]
\centering 
\includegraphics[width=5cm]{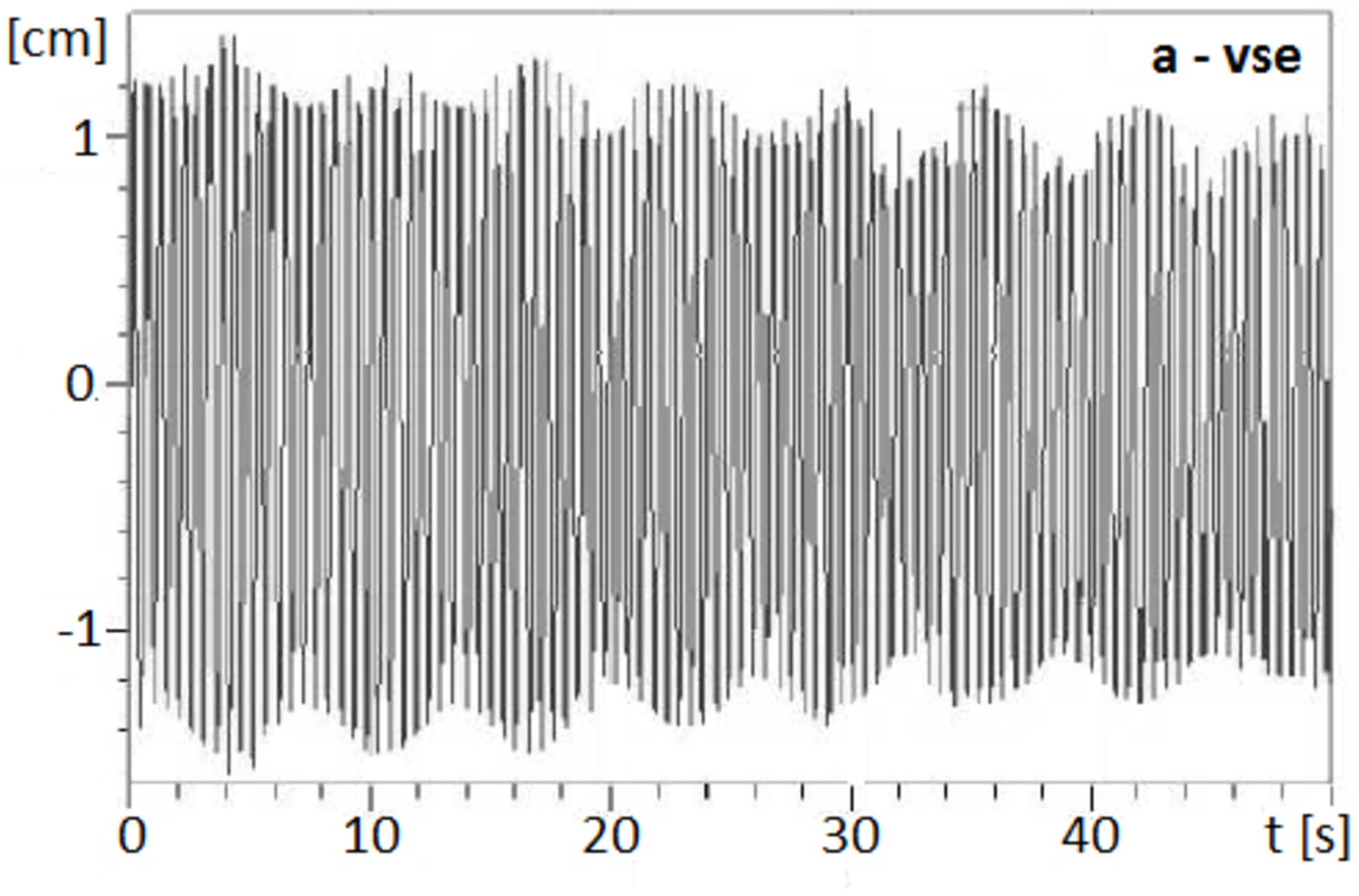}
\includegraphics[width=5cm]{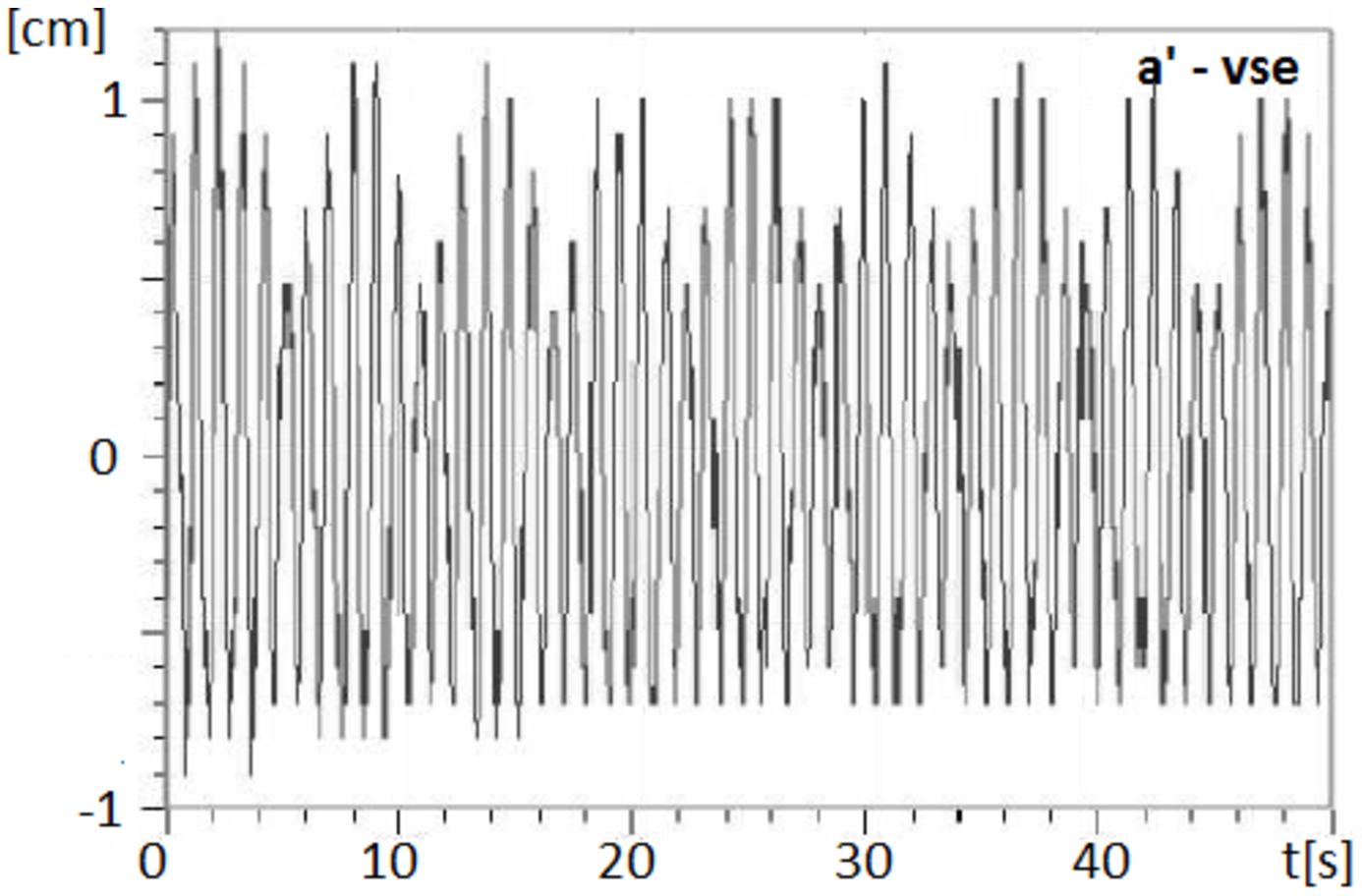}
\\
\includegraphics[width=5cm]{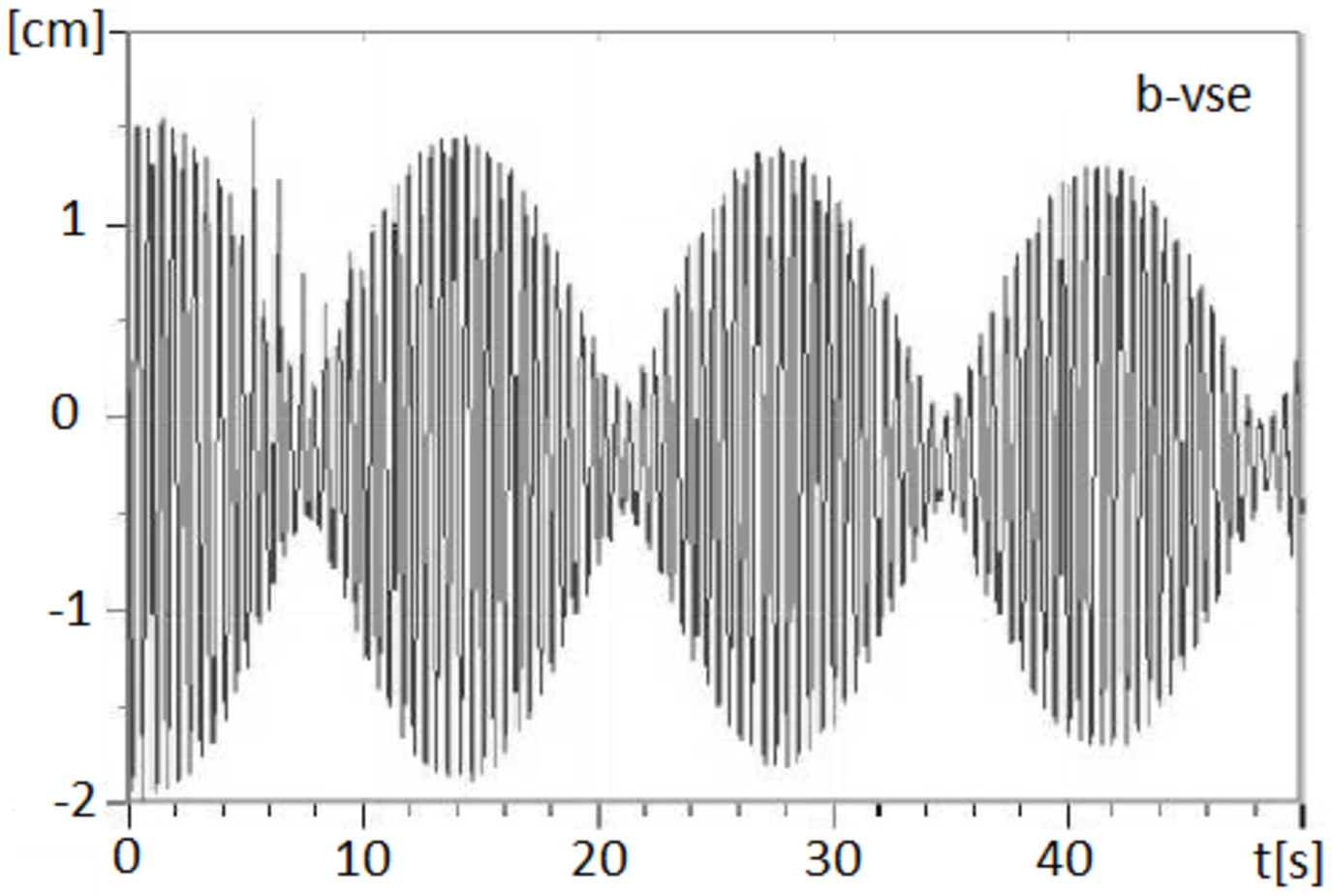}
\includegraphics[width=5cm]{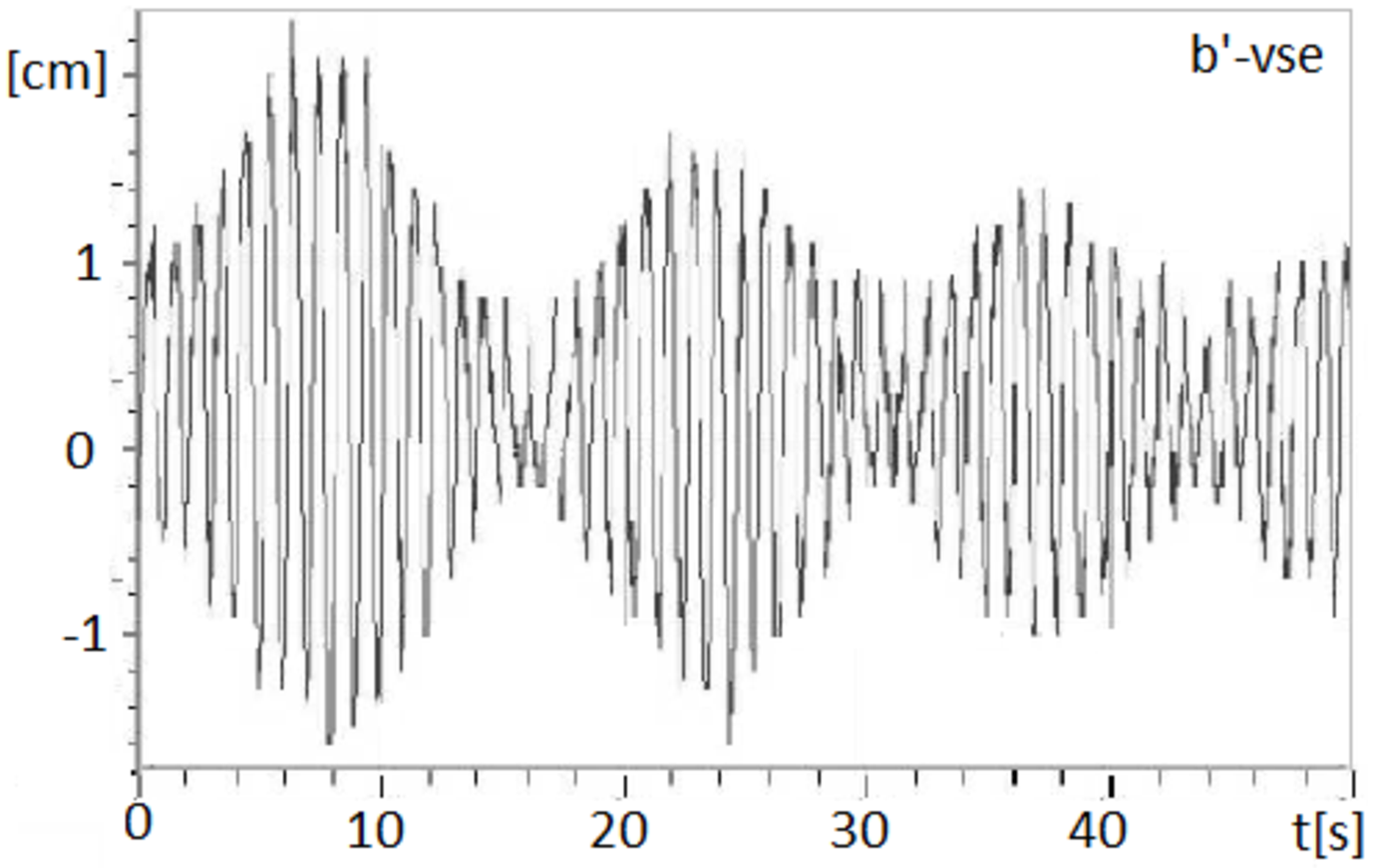}
\caption{\label{55g-son1+son2} The Figs. a, a', b and b' refer respectively to FR=1.85 (out of resonance) and FR=1.92 (not far from resonance). They are obtained with vse excitation. The envelope modulation of waveform1 and waveform2 in Figs. a and a' indicates that some energy exchange between the two system modes takes place. Figs. b and b' show that the system is unstable, there is a complete motion interchange between the two modes, because of this the lobes of the modes are in anti-phase.}
\end{figure}

\begin{figure}[!ht]
\centering
\includegraphics[width=5cm]{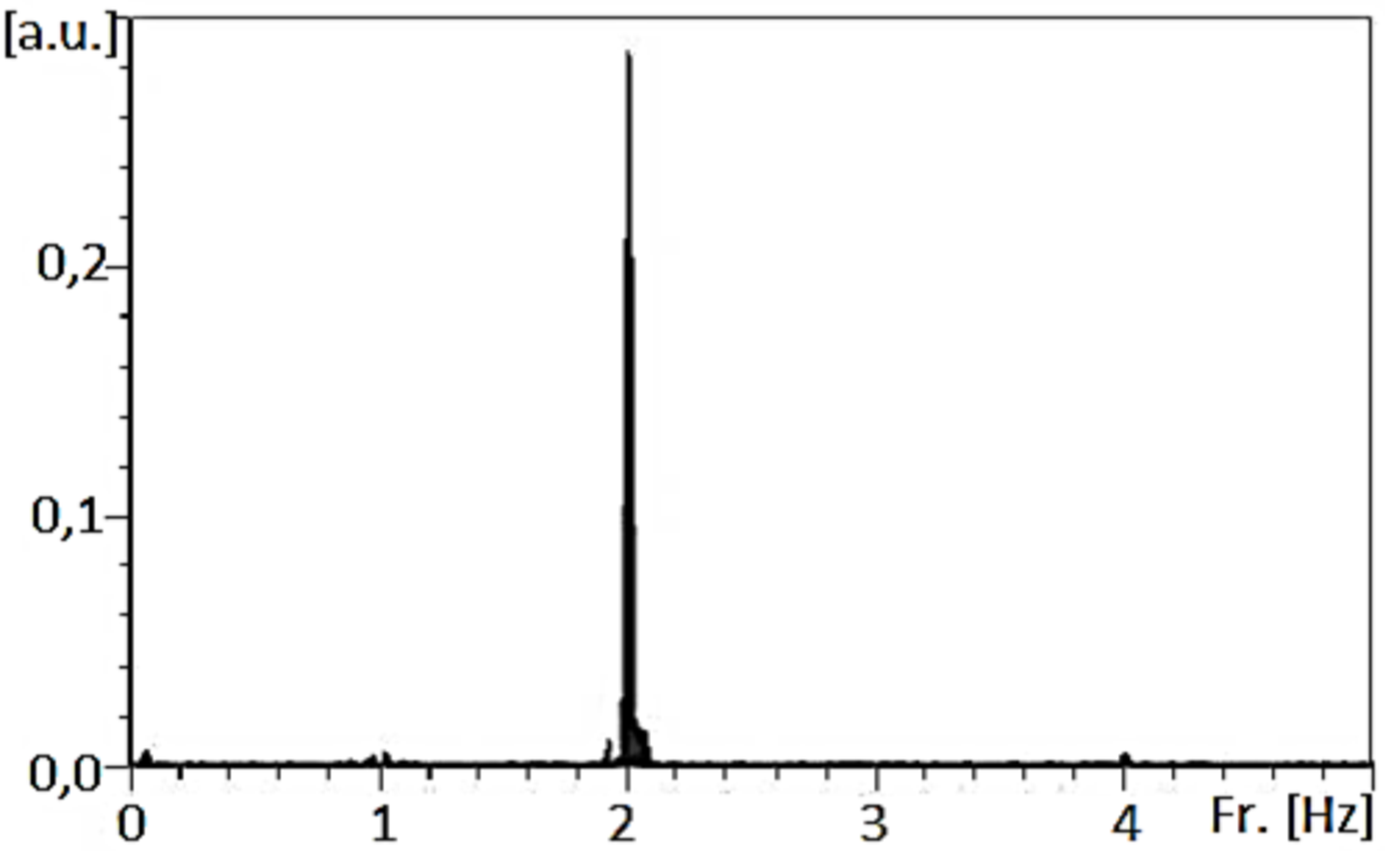}
\includegraphics[width=5cm]{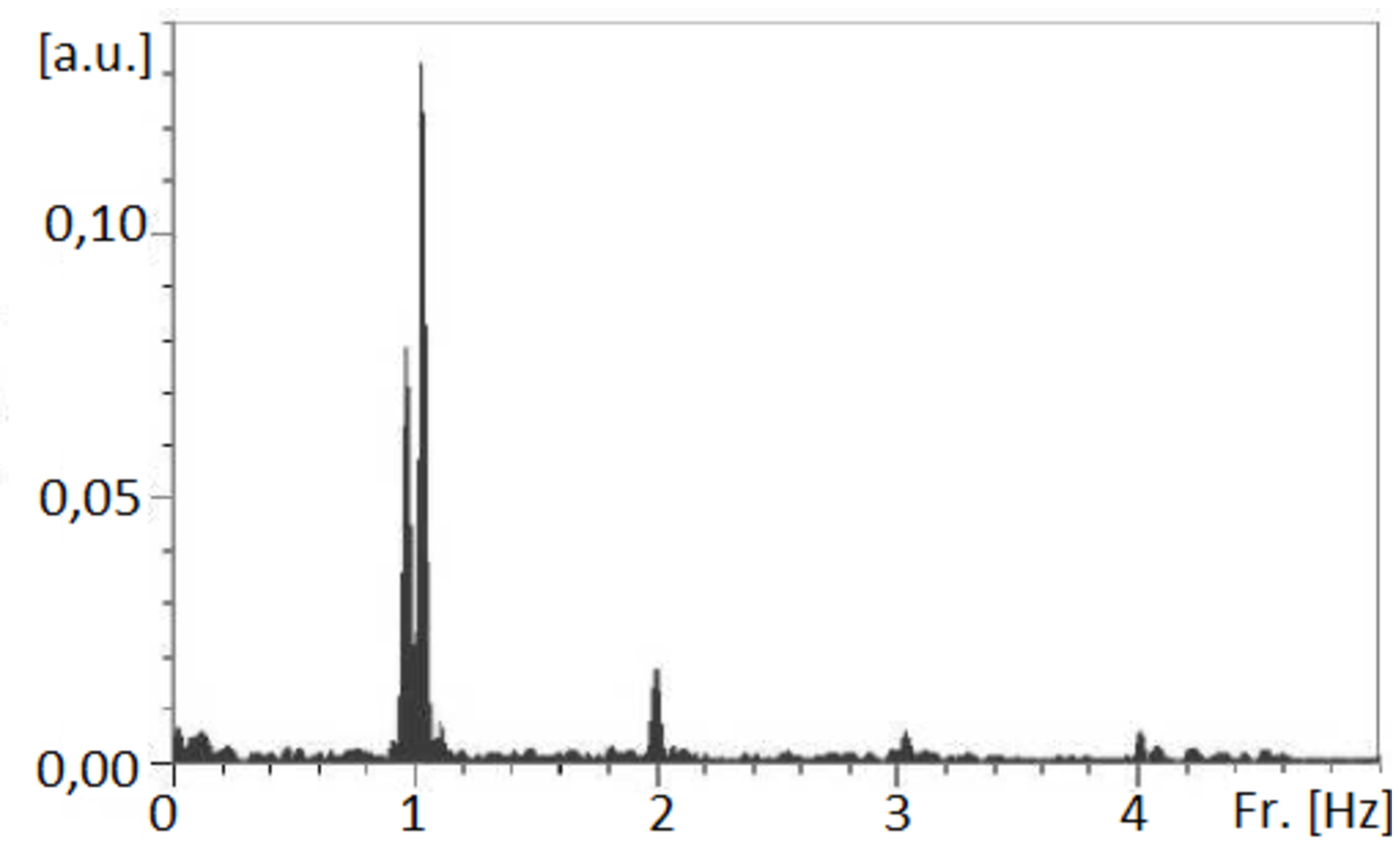}
\caption{\label{50-vse-fft-son1+2} The 50g Spectrum1 in the left frame shows the expected lines: the fundamental of the spring oscillation ($\nu_k$=2.03 Hz) and of the pendulum oscillation ($\nu_p$=1.06 Hz, quite faint) plus the very pronounced second harmonic 2$\nu_p$. Spectrum2 shows the fundamental pendulum line at 1.06 Hz, the clearly visible $\nu_k$=2.03
  Hz, and the pronounced idler frequency line at 0.97 Hz.}
\end{figure}

Waveform1 and waveform2, relative to the two configurations having FR equal to 1.85 and 1.92 (obtained by attaching masses of 55g and 50g respectively), are reported in Fig. \ref{55g-son1+son2}. The two spectra relative to the 50g waveforms are reported in Fig.
\ref{50-vse-fft-son1+2}. The other spectra are similar except for the fact that the frequencies are different. The vertical excitation was done by hand, elongating the spring by $\delta \ell \sim 1.5\, cm$. %


All the spectra1 have the main line $\nu_k$, a sign of the $\nu_p$ line, and a more evident harmonic $2\nu_p$ line. All the spectra2 have the two $\nu_p$ and $\nu_k$ lines as well as a pronounced idler line $\nu_i$. The relatively strong 2$\nu_p$ line in spectra1 (stronger than the $\nu_p$ lineline) is due to the fact that the vertical motion is modulated at that frequency by the pendulum motion via the term $mgcos\theta$ (left-right
invariant) and, in addition, by the fact that Sonar1 detects the pendulum oscillation in the peculiar way shown in Fig. \ref{view-sonars-moti}. The intensity of the idler line increases with the increase of the mode exchange.

The waveforms obtained by the configuration with FR=1.99 (45g, in resonance) were very similar to those with FR=1.92.
The same result is obtained with FR=2.07 (40g, not far from resonance). The heavier the masses, the lower the mode exchange and also the waveforms' modulation due to the decrease of the mode coupling. With this vse excitation, when a mass of 70g is used (FR=1.67), the mode exchange disappears. The resonance width results work out as $\Delta (FR) \sim 0.32$.

The direct motion observation provides information on the variation of the pendulum oscillation plane, the fraction of the energy exchange between the modes, and the shape of the mass trajectory. The trajectories take the shapes of a concave arc, a convex arc, and a simile Lissajous figure. Their amplitudes are within the space bounded by the two maximum convex and concave arcs, as shown in Fig. \ref{view-sonars-moti}.
  
The main findings from the waveforms analysis are as follows: the motion exchange between the two modes is complete within the FR=(1.9--2.1) interval; the higher the energy transfer fraction, the slower the motion transfer time from one mode to the other; the shape of the lobes is substantially exponential in accordance with the analytical calculation presented in reference \cite{cayton}, the beginning of waveform2 is often fuzzy since the relative phase between the two modes changes from undefined to defined.

The main findings from the analysis of the spectra are as follows: the input spring frequency $\omega_k$ is split into the two (pendulum) $\omega_p$ and (idler) $\omega_i$  frequencies when passing from vertical to pendular motion; the three frequencies satisfy the expected relation $\omega_k=\omega_p+\omega_i$; the three frequencies $\omega_k$, $\omega_p$, and $\omega_i$ vary by a few hundredths of a digit in repeated tests.

It can be inferred that the shift of the three frequencies is due to the auto-setting of the trajectory within the space enclosed by the concave-convex limiting arcs. The system, when choosing a particular trajectory, adopts a pendulum oscillation amplitude and an average spring length, that is - the $\omega_p$ and $\omega_k$ values. Our results and model are in line with those reported in Ref.\cite{olsson} which refers to the theoretical argumentation of Ref.\cite{minorsky} based on the method of the harmonic balance.

\begin{figure}[!ht]
\centering
\includegraphics[width=6cm]{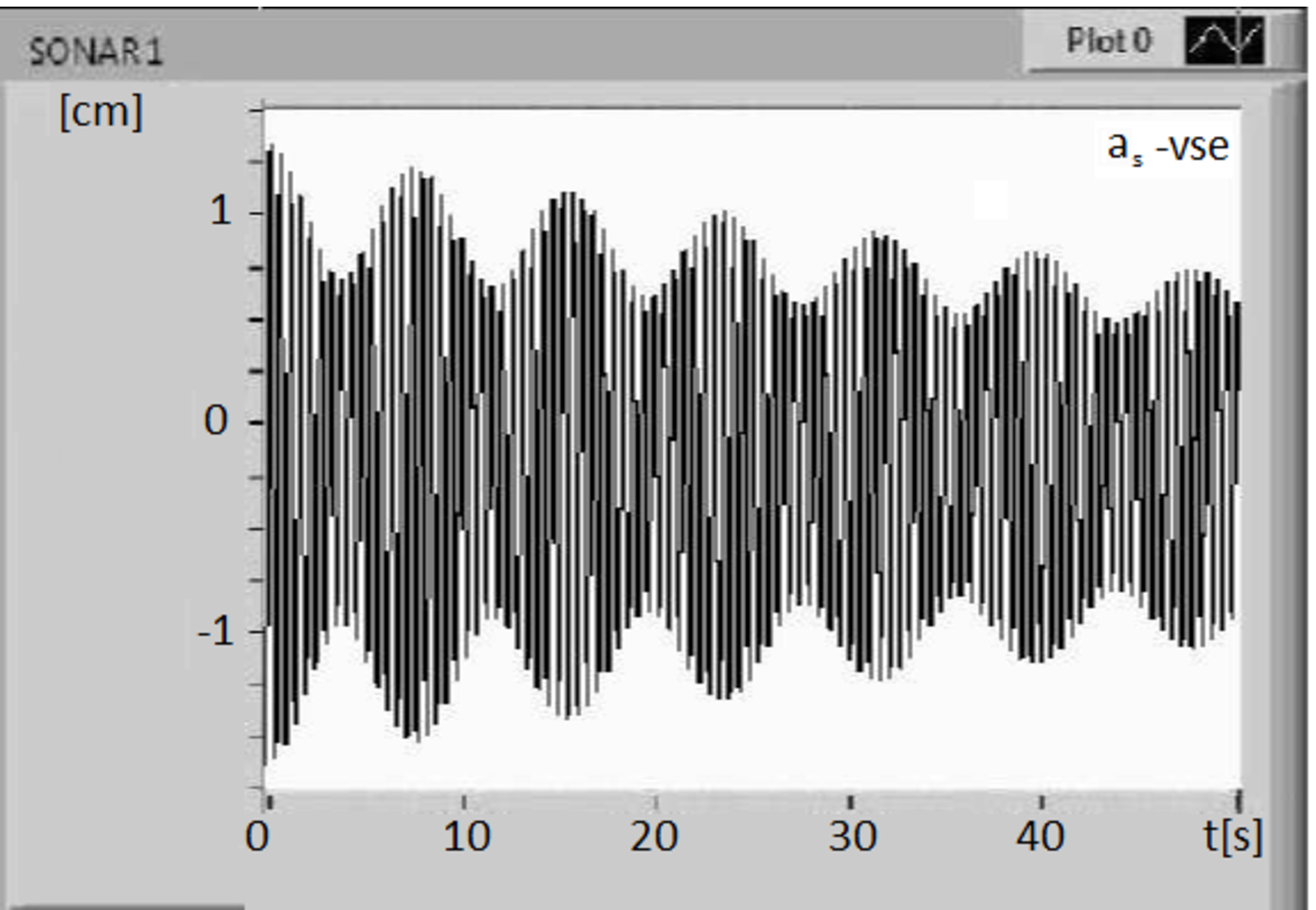}
\includegraphics[width=6cm]{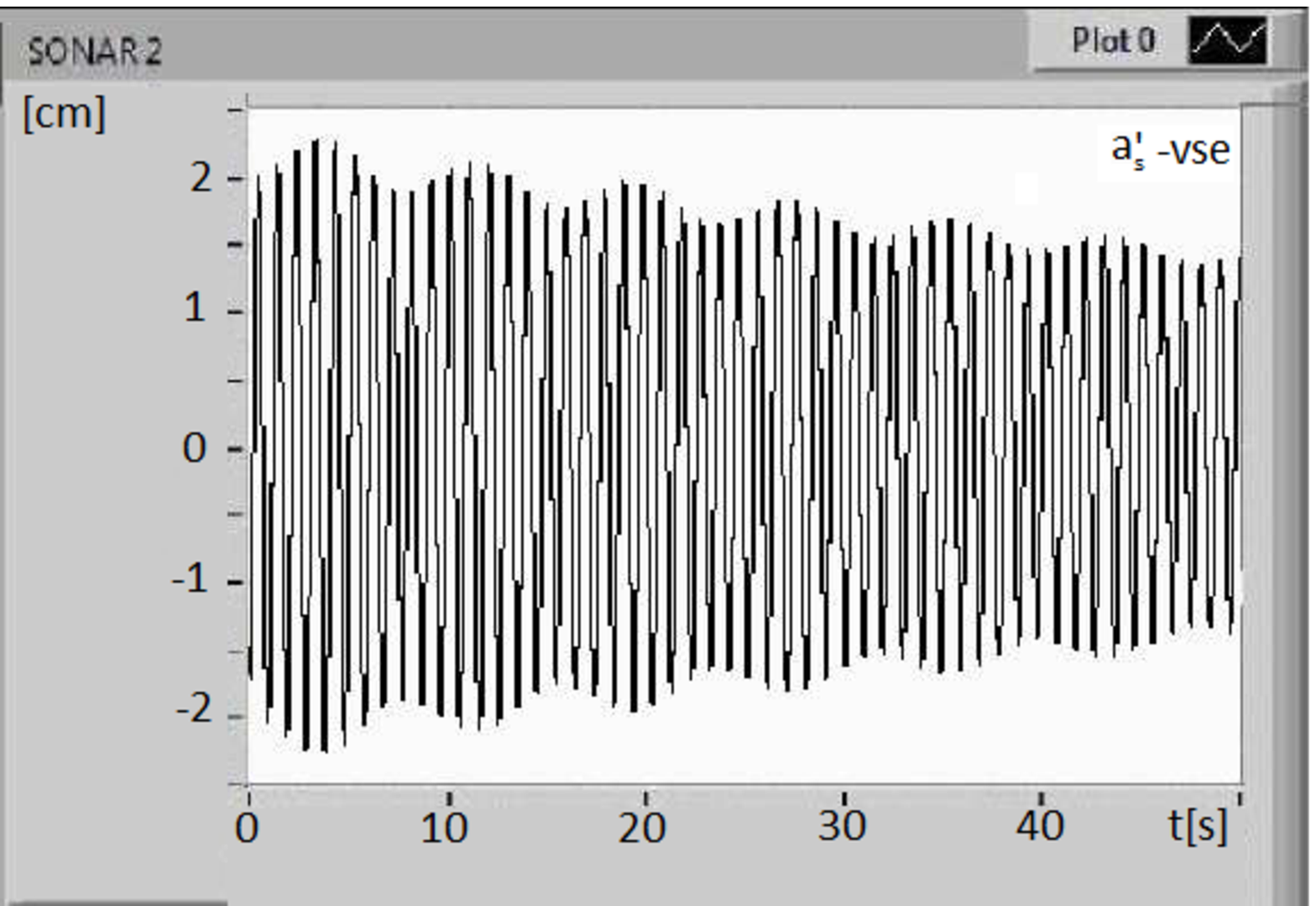}
\\
\includegraphics[width=6cm]{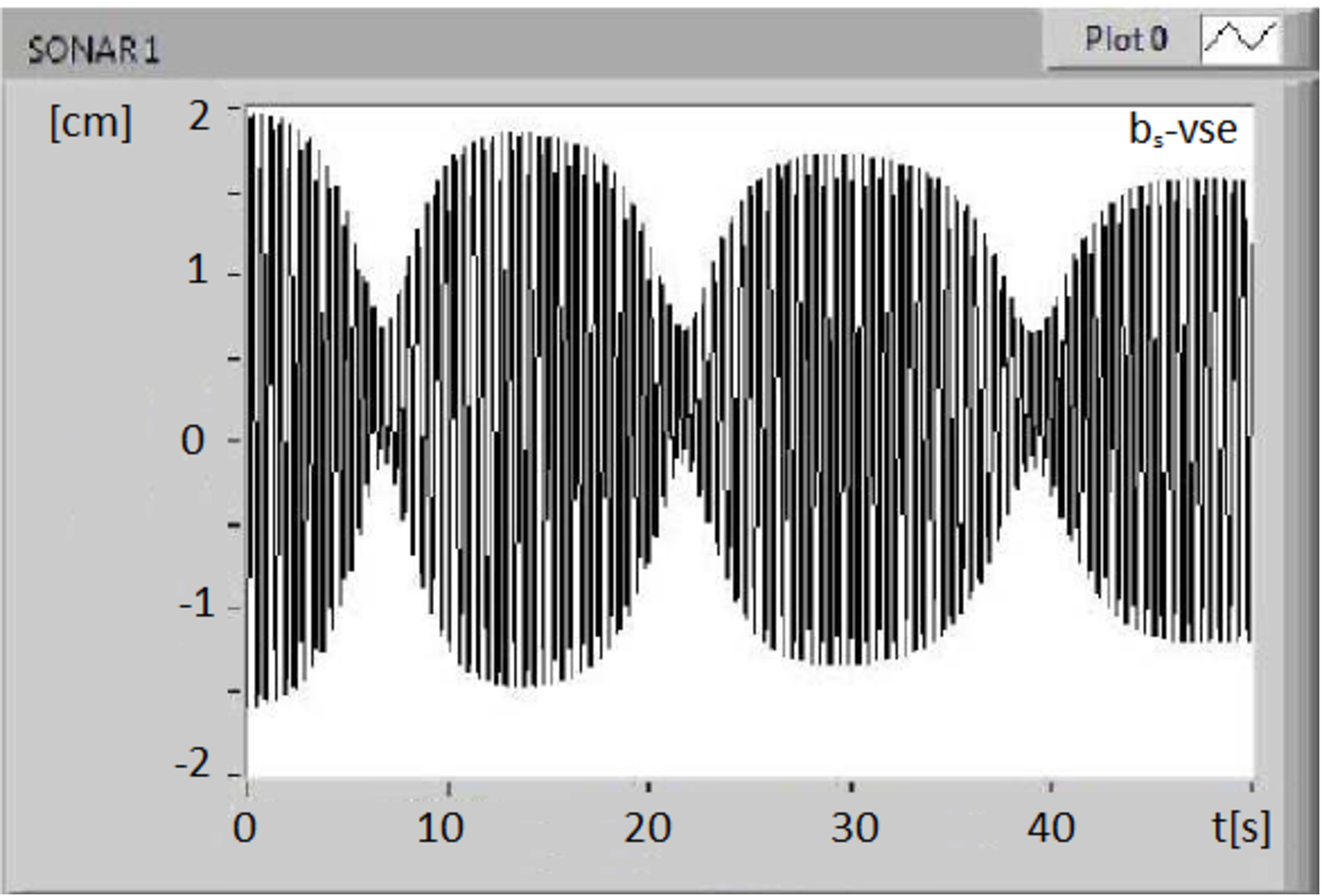}
\includegraphics[width=6cm]{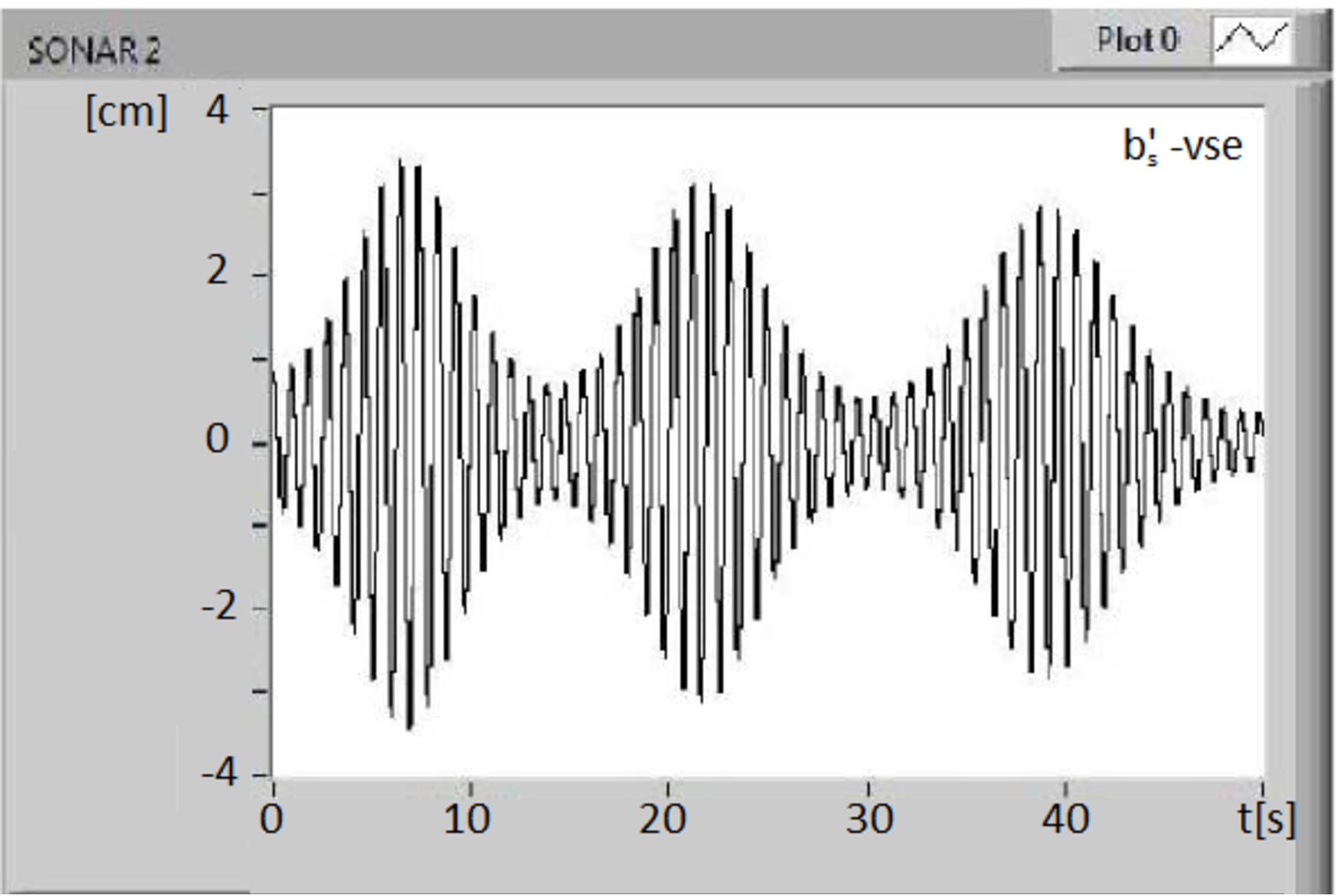}
\caption{\label{45g-simulz1+2} The simulations a, a' and b, b' refer respectively to FR = 1.85 and FR = 1.99, and are obtained with vse excitation. They largely reproduce the experimental results.}
\end{figure} 

The simulations relative to the configurations with masses of 45g and 55g, obtained with the theoretical Eqs. \ref{eq1-parametric} and \ref{eq2-parametric}, are presented in Fig. \ref{45g-simulz1+2}. They reproduce the experimental waveforms quite well. The simulation requires a lateral (even if very small) mass shift to onset the motion instability. In the actual system, spurious motions always turn on the instability. 


\subsection{Behavior with vle excitation}


In this section, the motion under the action of strong non-linearity is studied and compared with the previous one (vse). The strong non-linear regime is initiated by a strong excitation (vle). 

\begin{figure}[!ht]
\centering 
\includegraphics[width=5cm]{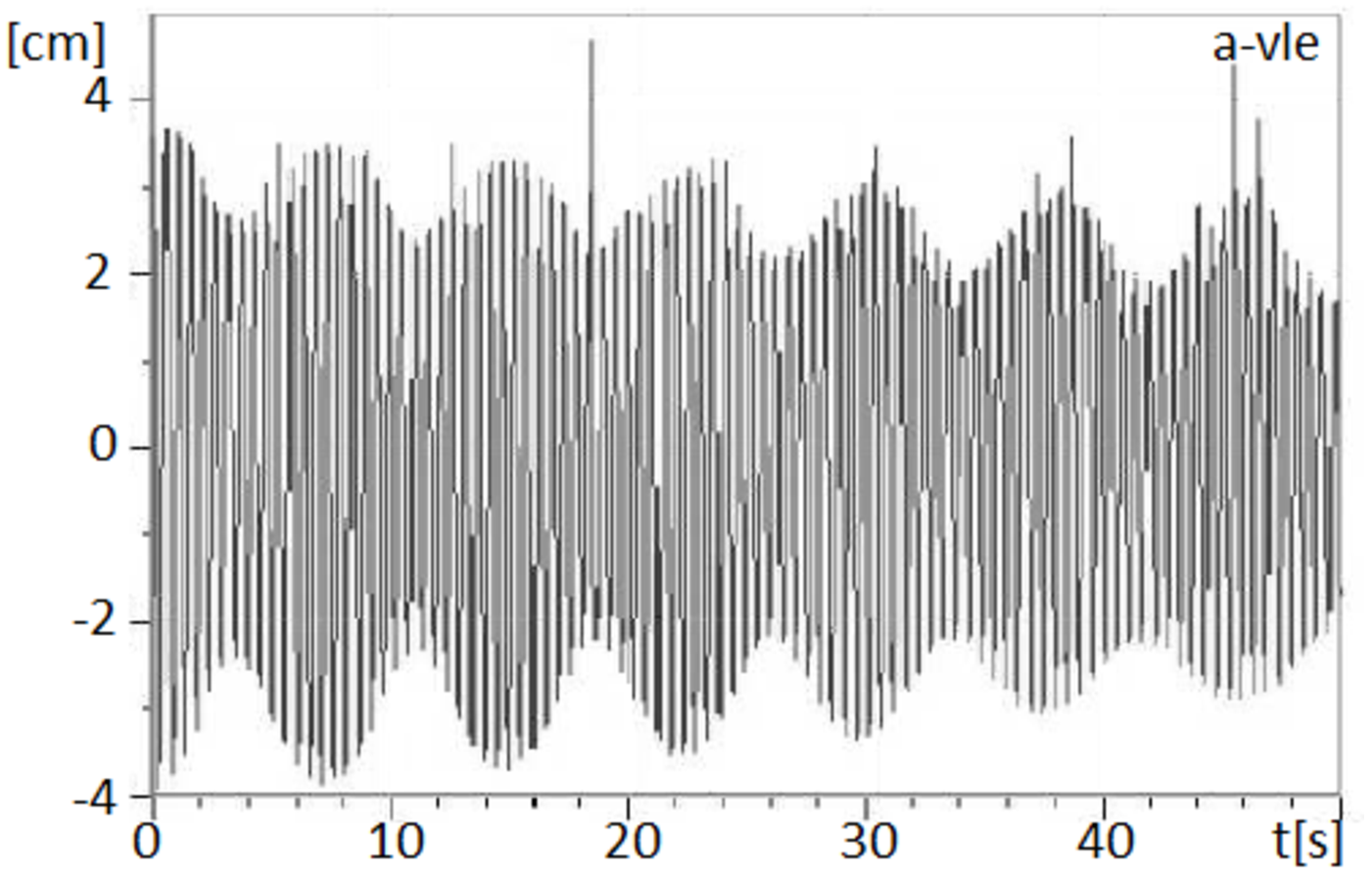}
\includegraphics[width=5cm]{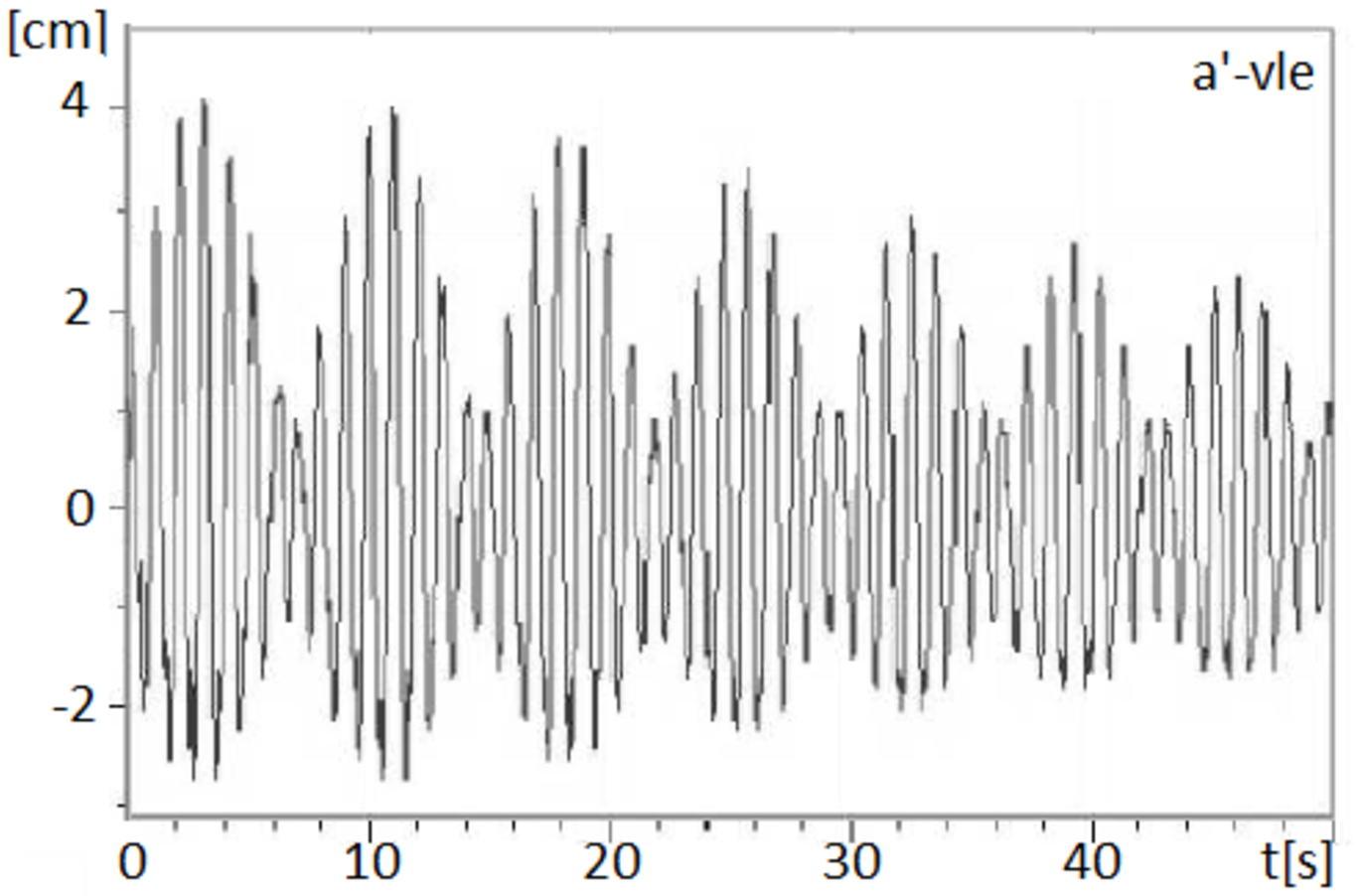}
\\
\includegraphics[width=5cm]{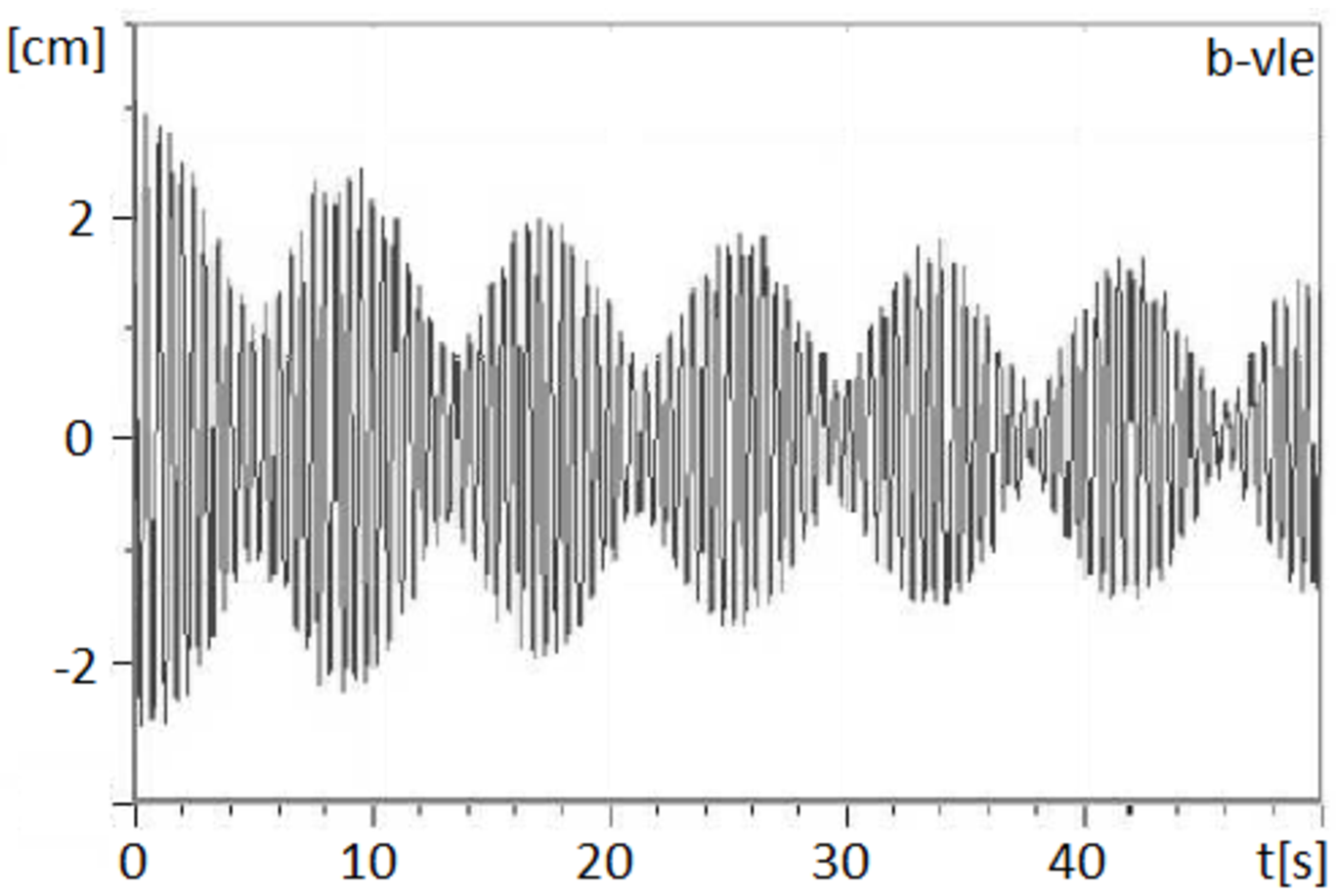}
\includegraphics[width=5cm]{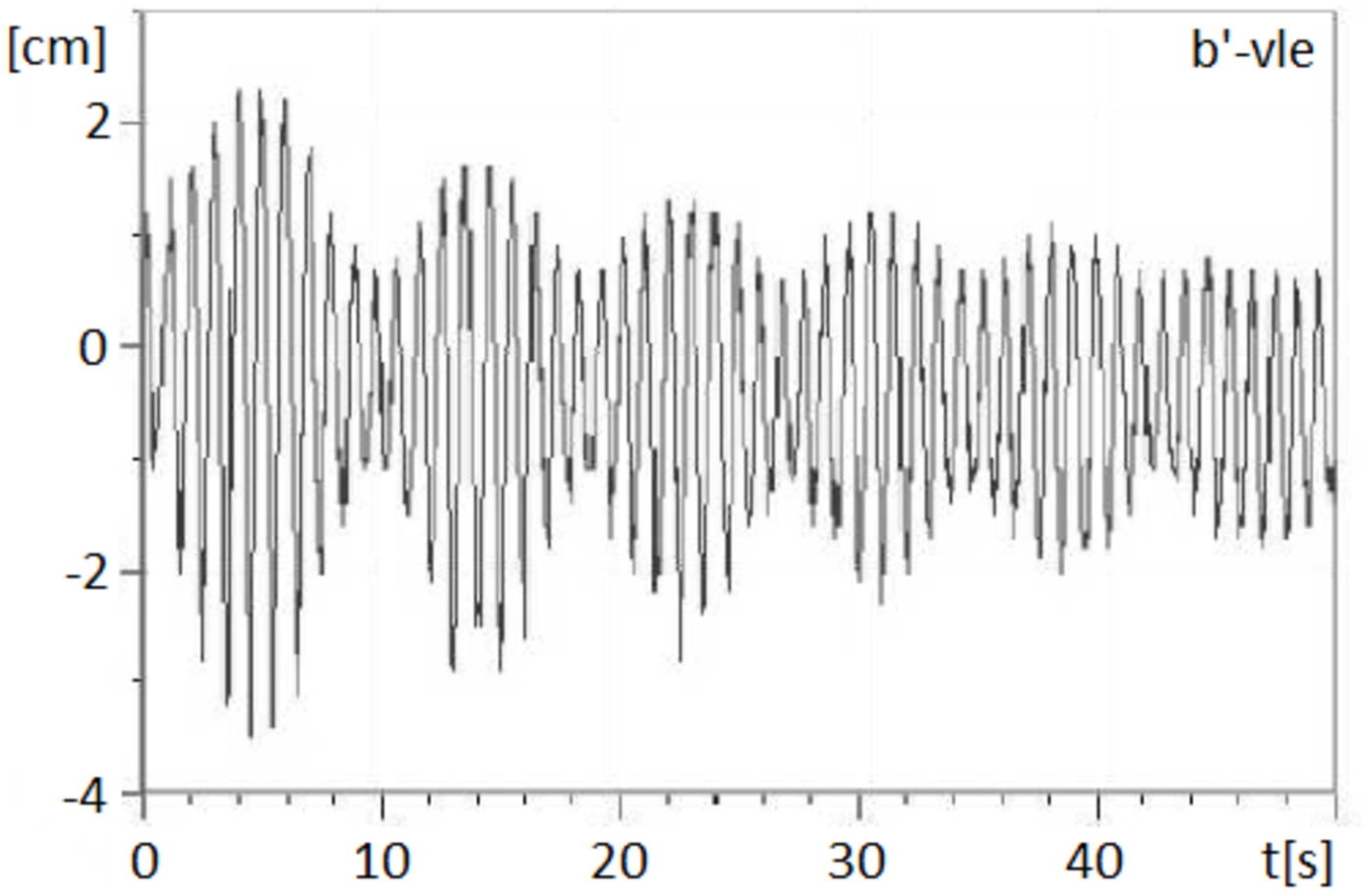}
\\
\caption{\label{55g-vle-waveforms1+2} The Figs. a, a', b and b' refer respectively to FR=1.85 and FR=1.92. They are
obtained with vle excitation. Vertical and transverse waveform1 and waveform2 (a, a') have a deep modulation, waveforms b and b' are completely modulated. The number of lobes here (with large
non-linearity) is higher than in the previous case (with small non-linearity). Therefore, the coupling strength increases with the growth of the oscillation amplitude.}
\end{figure}

The waveforms relative to the same 55g and 50g masses obtained with $\delta \ell \sim 3-4\, cm$ elongation are reported in Fig. \ref{55g-vle-waveforms1+2}.

\begin{figure}[!ht]
\centering 
\includegraphics[width=5cm]{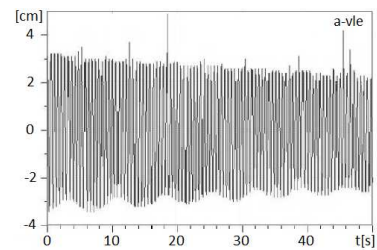}
\caption{\label{55g-vle-waveforms1-diretta} Waveform1, as detected by Sonar1, relative to the system configuration with FR=1.85 obtained with vle excitation.}
\end{figure}

By observing the motion directly, we can see that the trajectory ends up being a convex arc. The motion detection of this kind of trajectory by Sonar1, Figs.  \ref{view-sonars-moti} and \ref{Sonar1-pendolo}, leads to a waveform1 with an almost flat top edge, as shown by a typical example in Fig. \ref{55g-vle-waveforms1-diretta}. The shapes of the actual waveform1 are obtained by subtracting the wave superimposed by the detector, reported in Fig. \ref{Sonar1-pendolo}, reproducing so the shape of the waveforms observed with the naked eye.  Incidentally, the simulation helps to deal with this case.
    
The motion with vle excitation, compared to the previous vse excitation, is more disordered, the waveforms are more irregular, the mode exchange is faster, the resonance width is wider, and the spectra have additional lines and more pronounced idler lines. The increase of the non-linearity explains both the faster mode exchange and the wider resonance width, the latter leading to a more pronounced idler frequency  $\omega_i$ line. In fact, the observed mode exchange extends up to FR=1.6 (i.e. 80g mass). The additional frequencies present in the spectra are due to both the harmonics of large pendulum oscillation and entangled spring oscillation, and to the non-negligible spurious motions.

These results and considerations hold for masses which are both heavier than 55g (60g, 70g, 80g) and lighter than 50g (45g, 40g). The simulated waveforms reproduce all of the experimental ones to a large extent.

\textit{A note on frequency shifts}. The spring oscillation frequencies move towards lower values while the $\nu_p$ frequency remains stable (in passing from vse to vle excitation: in the case of 55g, from 1.95 to 1.92 Hz; in the case of 50g, from 2.03 to 2.01 Hz; and in the case of 45g, from 2.14 to 2.11 Hz). This merits consideration. This frequency shift moves further away from the resonant value FR, thus it can be said that the system moves further from resonance, while the observed parametric action is stronger. The two facts can coexist because the resonance width becomes larger. This result seems to indicate a reduction in the average value of the pendulum length. This is possible with a convex arch trajectory.


\subsection{Behavior with vertical small and large excitation plus a seed of transverse signal vse+s and vle+s}

In line with the idea that the system can behave as a frequency converter, a very small transverse shift was added to the vertical spring stretching. From this point of view, the non-linear interaction is the mechanism providing the energy transfer from pump to seed. In this case, there is still a resonance condition among the frequencies. It does not come from the mechanical system parameters, but from its capability of setting itself at the resonant condition of matched periodic motions.

The following experimental results support the conjecture: (I) the mode coupling extends itself to heavier masses, see the noticeable waveform1 modulation relative to the system configuration with FR=1.78 (60g mass) and the modulation even at a FR=1.6 (80g mass) configuration as shown in Fig. 11; (II) the complete mode exchange extends up to FR=1.84; (III) the system chooses a convex arc trajectory; and finally, (IV) the idler frequency line is neat. Simulations reproduce the measurements reasonably well. 

\begin{figure}[!ht]
\centering
\includegraphics[width=5cm]{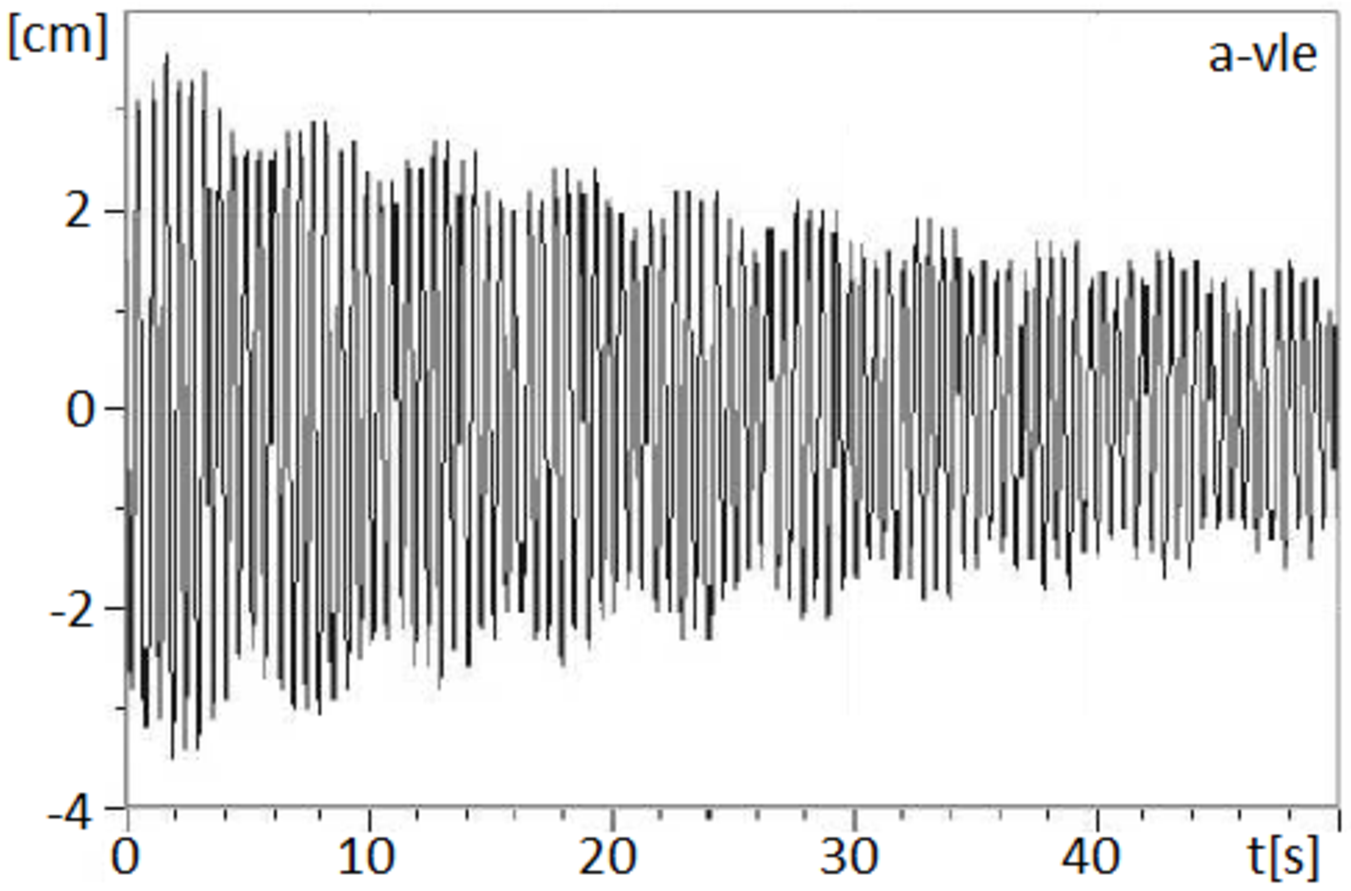}
\includegraphics[width=5cm]{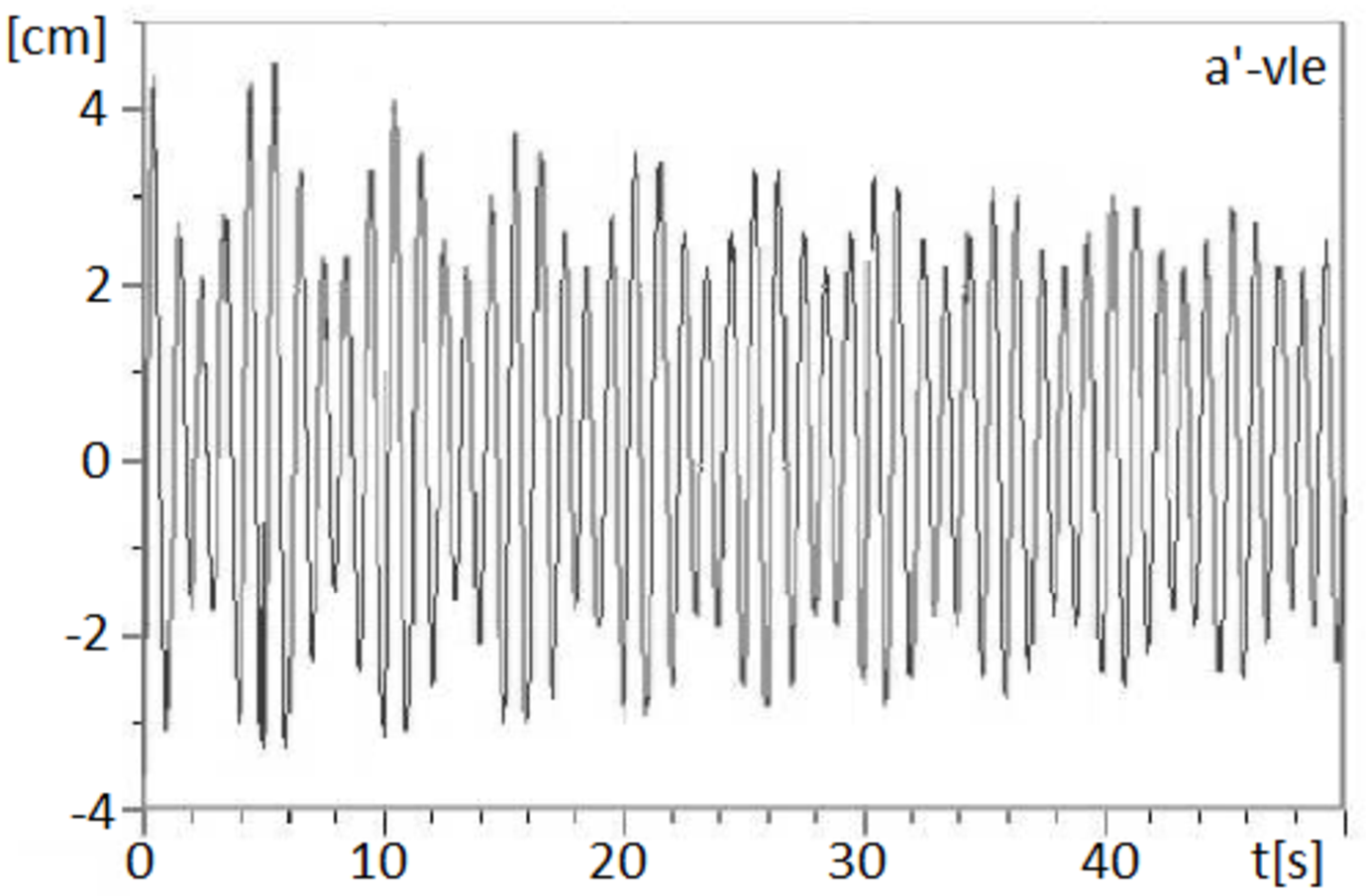}
\\
\includegraphics[width=5cm]{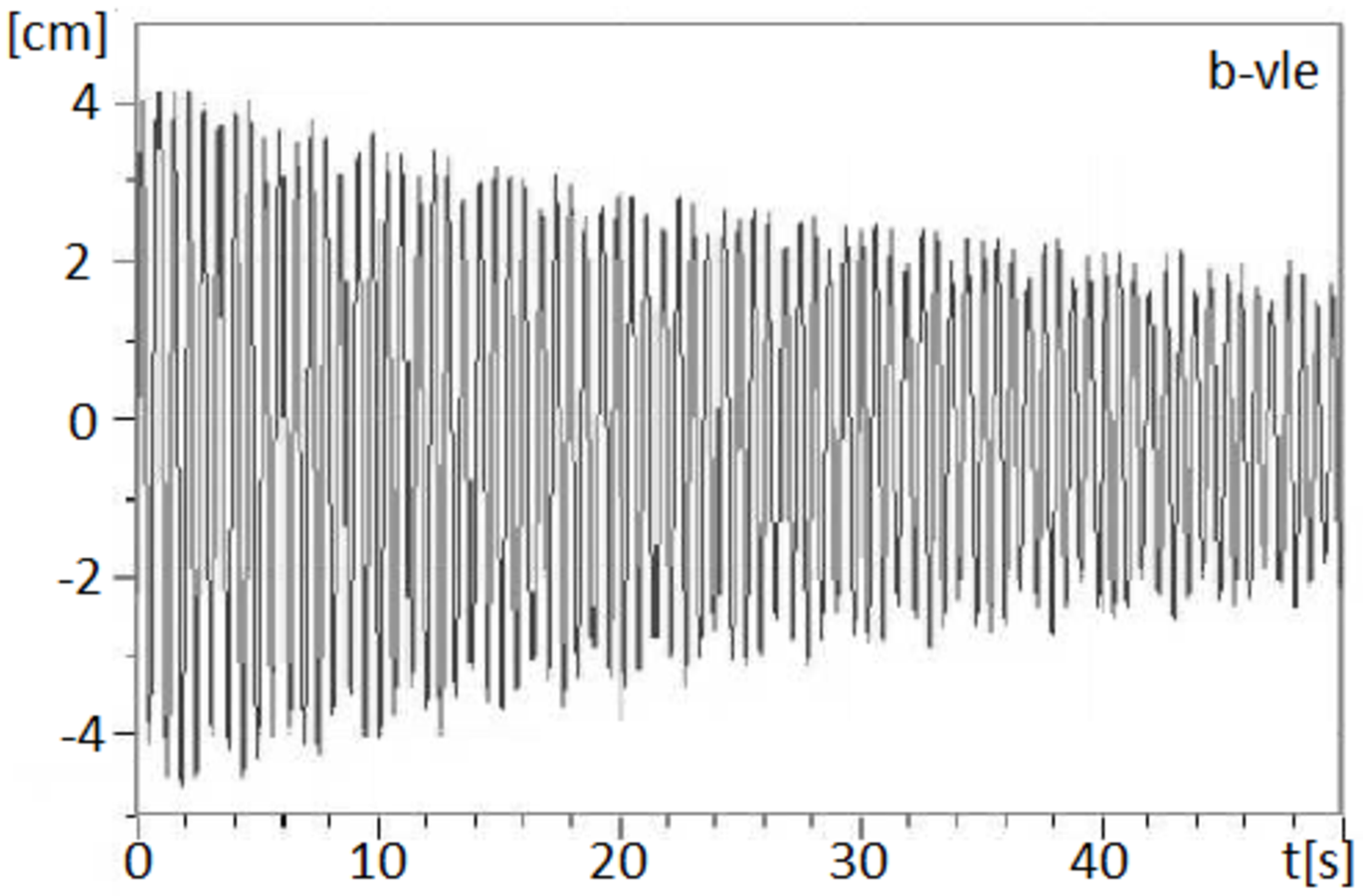}
\includegraphics[width=5cm]{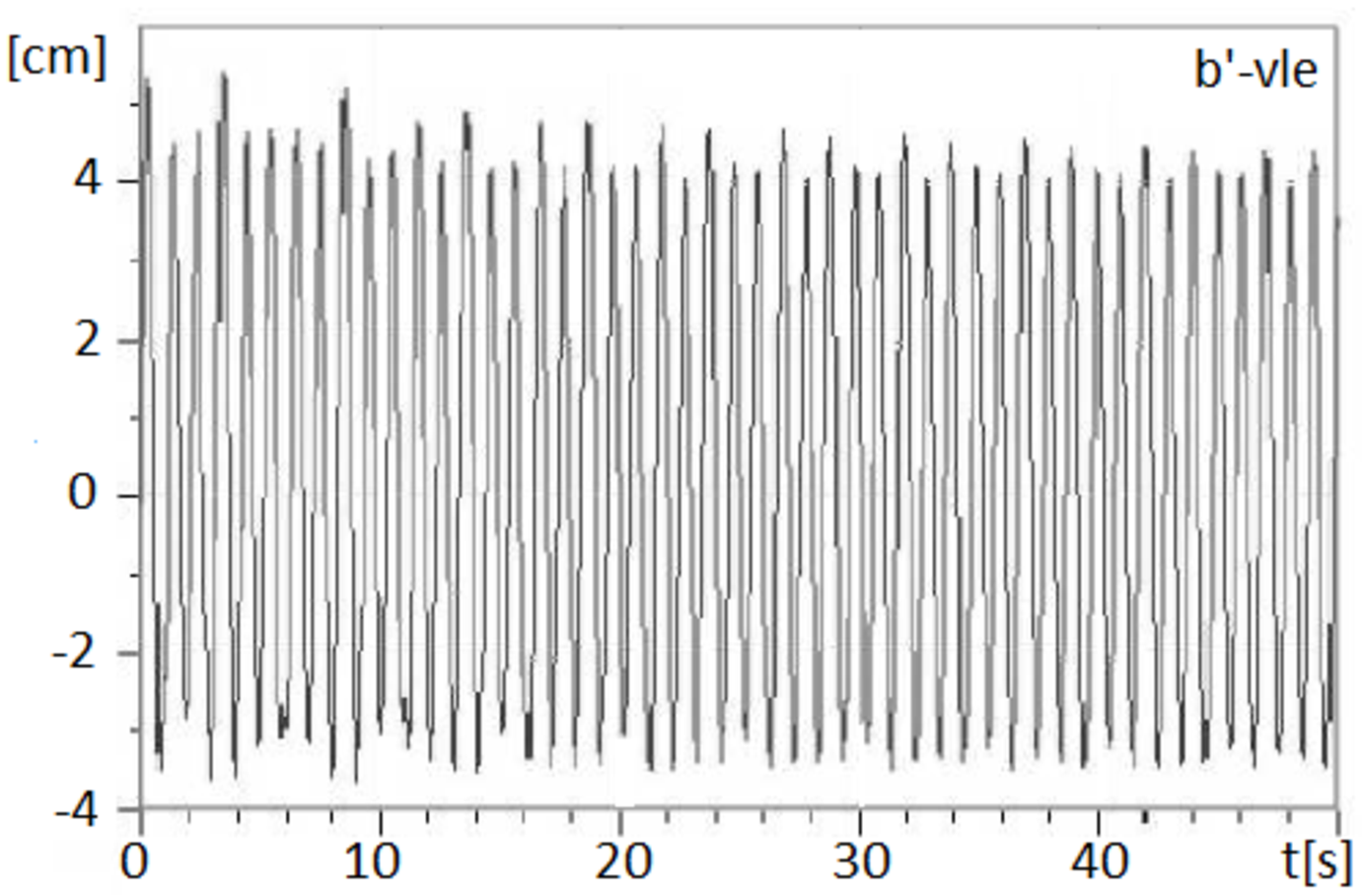}
\caption{\label{mle-60g} The Figs. a, a', b and b' refer respectively to FR = 1.78 and FR = 1.60. The evident
waveforms' modulation indicates the energy exchange between the two modes.}
\end{figure}

The start of the \textquotedblleft amplified\textquotedblright \, oscillation is neat. This is explained by the fact that the initial relative phase between the two modes is $\pi$, as was theoretically calculated in Ref. \cite{cayton}. The setting of an initial $x(0) \neq 0$ causes the onset of the exchange between the two modes, as predicted by the coupling term RHS of Eqs.~\ref{eq1-parametric} and \ref{eq2-parametric}.


\subsection{Behavior with small and large transverse excitation tse and tle}

Transverse excitation is intrinsically different from vertical excitation because the pendulum oscillation of the mass generates an oscillation of the gravitational force on the spring (by the term $mg \cos\theta$), and this in turn generates a vertical mass oscillation. In short, this excitation couples the two modes directly, and the mode exchange becomes inherent in the system. This fact leads to experimental results with neat waveforms and relative onsets, and also to neat spectra showing a clear idler frequency line. Overall, the system behavior is similar to the previous one with vertical excitation, although some differences can be observed. The resonance width obtained is a little larger, the pendular oscillation does not transfer completely to the vertical oscillation even at resonance, and the 2$\nu_p$ line in spectrum1 is so strong that it becomes comparable to the $\nu_k$ line.

The presence of the $\nu_k$ and 2$\nu_p$ frequencies in the vertical motion spectrum indicates that the system arranges itself into such a motion configuration as to allows mode coupling, i.e. a configuration in which the periods of the two vertical and transverse oscillations are close to resonance. In fact, the frequencies of the two oscillations move with respect to previous values.


\section{Discussion and conclusion}

A vertical spring--mass system can oscillate both vertically and laterally. The two motions are non-linearly coupled. When its parameters, viz., length, spring constant and mass, lead to oscillation frequencies in resonance, the motion is composed of sideways swinging and vertical spring oscillation, and the two oscillations interchange with each other.  
The spring--mass system exhibits a mechanical peculiarity: it is possible for mass trajectories to be found all around the space delimited by the concave and convex arcs.
Mostly, concave and convex trajectories are observed, but a type of simile Lissajous curve is also sometimes seen. The motion trajectory depends on the kind of excitation. These trajectories are also traced by motion simulations with different excitations. The following main features are observed: (1) {\em strong mode exchange}; (2) {\em large resonance width}; (3) {\em pendular and spring oscillation frequencies covering a certain interval}; and finally (4) {\em settlement of the mass trajectory within a concave-convex arc space}. The increase in the vertical stretching increases both the resonance width and the velocity of the mode exchange; the spring frequency $\omega_k$ is transformed into the pendulum frequency $\omega_p$ plus the idler frequency $\omega_i$ satisfying the relation $\omega_k = \omega_p + \omega_i$; these three frequencies vary by a few hundredths of a digit in repeated tests.

All these measured and observed features lead us to conjecture that the non-linear system adapts itself to a mechanical configuration suitable for mode interaction. The ability of the system to end up with configurations with a frequency ratio FR inside the resonance curve width depends on the fact that the amplitudes of the two oscillations and the spring elongations (thus the two motion frequencies) adapt themselves for the exchange between the two motions. In other words, the mass trajectory can arrange itself within a wide space and, in addition, with different shapes. The oscillation frequencies of the two motions are entrained by the parametric action, the frequency of the pendular oscillation follows the string elongation, and the frequency of the spring oscillation is in turn dragged by the oscillation of the gravitational force component. This aspect of the motion conformation to the parametric interaction is also supported by the observation of many lines in the waveform spectra relative to complex trajectories. It can be inferred that the system accomplishes the adaptation of the oscillation periods by developing motion harmonics. For completeness, it could be added that the idea that the spurious motions present in the actual spring--mass system foster the parametric behavior.

The system adaptability to different motion configurations plus the fact that the pendulum motion causes a variation in the spring elongation by the appended mass (i.e. energy is exchanged between the two modes independently of the spring and pendulum frequency ratio) supports the view of the system as a parametric converter.

The parametric non-linear interaction between the two spring and swinging oscillations can be modeled in the same way as a three-wave non-linear parametric interaction in a plasma or a non-linear crystal. In this model, the input pump wave is converted into the output up- or down-wave and a third wave, called the idler wave. Within this framework, we can associate the vertical oscillation to the pump wave and the small side shift to the seed used for the frequency conversion.

The motion equations developed for this systematic study of the parametric behavior of the spring--mass system largely reproduce all of the experimental observations. They proved very useful in analyzing the many sophisticated aspects of the complex motion.

The system is used to study harmonic oscillator physics in a first-year laboratory course. In a laboratory organized so that students can assemble their own apparatus as they see fit (picking components from a pile), the spring--mass system is not likely to reproduce ideal harmonic motion. There is a high probability that they will set-up a system which shows complex dynamics. Furthermore, when testing the spring--mass physics law $\omega_k^2=k/m$, students have to apply different masses, therefore often run into parametric instability. Hence, studying the spring--mass system in a laboratory where students pick their own parts addresses both harmonic and parametric behavior. The spring--mass experiment can be treated in class as a harmonic oscillator and then developed towards a parametric oscillator. This teaching sequence allows students to approach the challenging, complex physics content of parametric behavior.

The rather complex -- but manageable -- experiment and the gap between the ideal theoretical case and the actual experimental case make this experiment a useful tool for stimulating an aspiring physicist's thinking.


\section{Acknowledgments}

Special acknowledgments to M. Michelini for supporting this research and for her critical and useful reading of the manuscript. We gratefully acknowledge technical assistance from D. Cipriani and S. La Torre.


\end{document}